\documentclass[fleqn,usenatbib]{mnras}

\usepackage{newtxtext,newtxmath}

\usepackage[T1]{fontenc}
\usepackage{ae,aecompl}

\usepackage{hyperref}
\usepackage{natbib}
\usepackage{amsmath}
\usepackage{footnotebackref}
\usepackage{empheq}
\usepackage{amsmath,amsfonts,amssymb}
\usepackage{graphicx}
\usepackage{wasysym}
\usepackage{chngpage}
\usepackage{array}

\usepackage{multicol}        
\usepackage{bm}		
\usepackage{pdflscape}	

\newcommand{\kepler}{\textit{Kepler}}
\graphicspath{{./}{figures/}}

\title{Dynamical instability and its implications for planetary system architecture}

\author[Wu et al.]{
Dong-Hong Wu$^{1,2}$,
Rachel C. Zhang$^{3}$,
Ji-Lin Zhou$^{1}$\thanks{zhoujl@nju.edu.cn},
Jason H. Steffen$^{2}$\thanks{jason.steffen@unlv.edu}
\\
$^{1}$School of Astronomy and Space Science and Key Laboratory of Modern Astronomy and Astrophysics in Ministry of Education,\\
Nanjing University, Nanjing 210093, China\\
$^{2}$University of Nevada, Las Vegas, Department of Physics and Astronomy, 4505 S Maryland Pkwy, Box 454002, Las Vegas, \\
NV 89154, USA\\
$^{3}$Massachusetts Institute of Technology, 77 Massachusetts Ave, Cambridge, MA, 02139, USA
}

\pubyear{2018}

\begin{document}
\label{firstpage}
\pagerange{\pageref{firstpage}--\pageref{lastpage}}
\maketitle

\begin{abstract}
We examine the effects that dynamical instability has on shaping the orbital properties of exoplanetary systems.  Using N-body simulations of non-EMS (Equal Mutual Separation), multi-planet systems we find that the lower limit of the instability timescale $t$ is determined by the minimal mutual separation $K_{\rm min}$ in units of the mutual Hill radius. Planetary systems showing instability generally include planet pairs with period ratio $<1.33$.  Our final period ratio distribution of all adjacent planet pairs shows dip-peak structures near first-order mean motion resonances similar to those observed in the \kepler\ planetary data.  Then we compare the probability density function (PDF) of the de-biased \kepler\ period ratios with those in our simulations and find a lack of planet pairs with period ratio $> 2.1$ in the observations---possibly caused either by inward migration before the dissipation of the disk or by planet pairs not forming with period ratios $> 2.1$ with the same frequency they do with smaller period ratios. By comparing the PDF of the period ratio between simulation and observation, we obtain an upper limit of 0.03 on the scale parameter of the Rayleigh distributed eccentricities when the gas disk dissipated.  Finally, our results suggest that a viable definition for a ``packed'' or ``compact'' planetary system be one that has at least one planet pair with a period ratio less than 1.33.  This criterion would imply that 4\% of the \kepler\ systems (or 6\% of the systems with more than two planets) are compact.

\end{abstract}


\begin{keywords}
planets and satellites: dynamical evolution and stability -- methods: numerical
\end{keywords}



\section{Introduction}

\begin{figure}
\vspace{0cm}\hspace{0cm}
\centering
\includegraphics[width=\columnwidth]{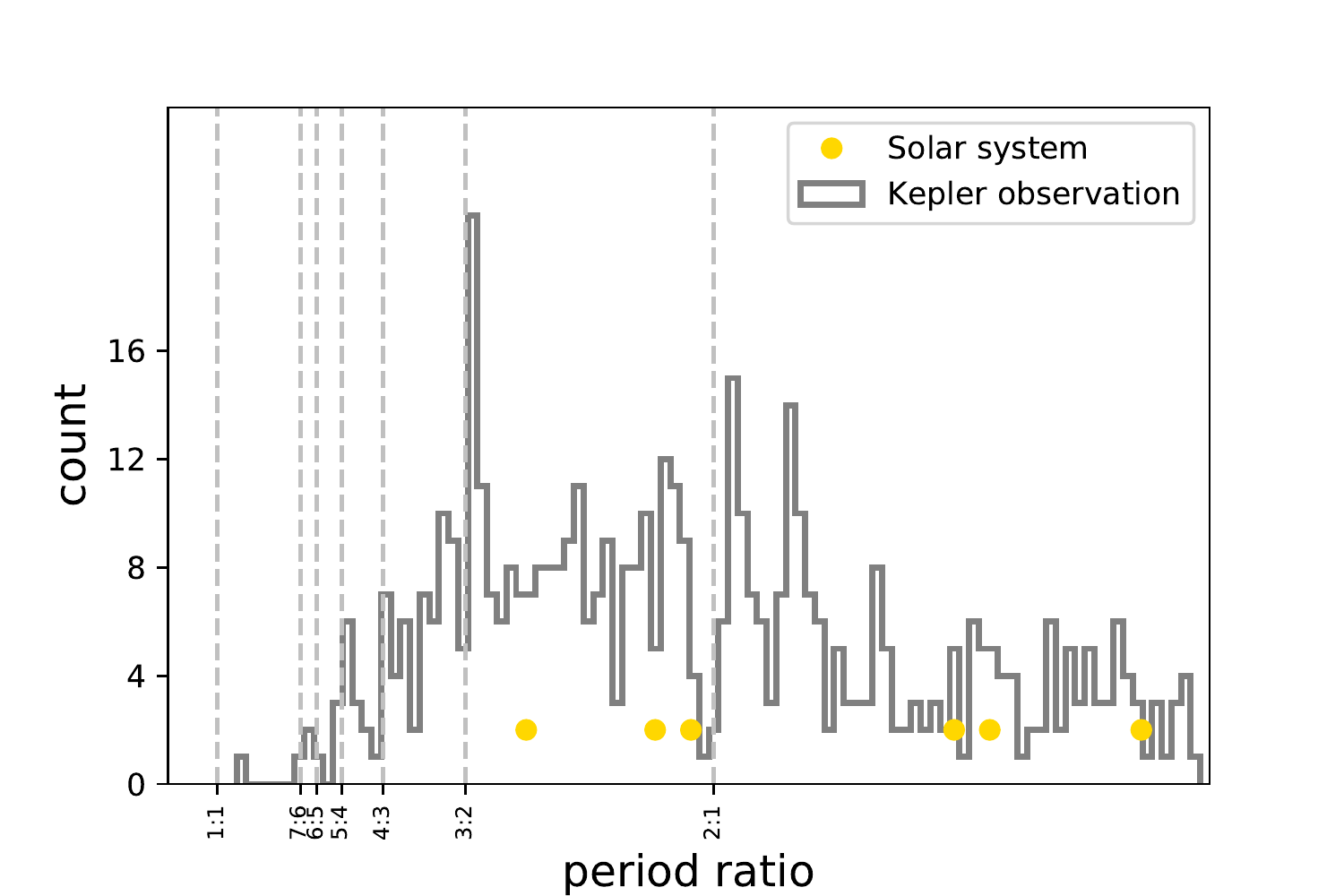}
\caption{Period ratio distribution of the \kepler\ adjacent planet pairs.  The samples are obtained from Q1-Q17 DR25 of NASA Exoplanet Archive and only confirmed planets are shown here.  The yellow dots are planets in the solar system.}
\label{observation}
\end{figure}

About $40\%$ of planets discovered by the \kepler\ spacecraft are in multi-planet systems \footnote{https://exoplanetarchive.ipac.caltech.edu}, some of which have small orbital period ratios between neighboring planets.  The observed period ratios between adjacent pairs (Figure \ref{observation}) show that most of the period ratios are smaller than three, and there is a pile-up of period ratios around the 3:2 and 2:1 mean motion resonances (MMRs).  The existence of planet pairs near first-order MMRs is often ascribed to disk migration \citep{snellgrove2001,lee2002,lee2009,wang2014}.  However, we might expect the overabundance to be larger if disk migration is common, though \citet{Pan:2017} suggested that resonance capture is more difficult for smaller planets in a disk.  Additionally, more planet pairs are observed on the far side of MMRs rather than being symmetrically distributed around them \citep{lissauer2011,fabrycky2014}.  Numerous mechanisms have been proposed to explain the asymmetrical period ratio distribution around MMRs including dissipative resonant repulsion \citep{lithwick2012,batygin2013}, stochastic and smooth migration \citep{rein2012a}, interactions between the planets and the planetesimal disk \citep{ford2015}, in-situ growth of planets \citep{petrovich2013}, and  planet-planet interactions \citep{pu2015}.

Most of the \kepler\ planetary systems are perceived as being quite compact, often containing multiple planets with orbital periods shorter than Mercury.  However, since the dynamics of the systems are generally scale invariant (dictated primarily by orbital period ratios rather than the orbital periods themselves \citep{Rice2018}) the term ``compact'' is ambiguous.  For example, compared with the physical size of the orbits of \kepler\ planets, the planets in our solar system are relatively far apart.  However, they have similar period ratios---the quantity that is more fundamental---to those observed in \kepler\ planet pairs (shown in Figure \ref{observation}).  Thus, either the \kepler\ planetary systems are less compact or the solar system is more compact than commonly envisioned.

The relationship between the spacing of the planets in a system and the stability of that system has been studied extensively.  For two planet systems, there are several stability criteria including Hill stability \citep{marchal1982,gladman1993,veras2004}, Lagrange stability \citep{barnes2006,zhou2003,barnes2007}, and the resonance overlap criterion \citep{wisdom1980,duncan1989,mardling2008,deck2013,ramos2015,Hadden:2018}.  For planetary systems that include more than two planets, the dynamics becomes more complex.  \citet{quillen2011} studied three-body resonance overlap in closely-spaced multi-planet systems.  Most other results, however, are based on numerical simulations \citep{chambers1996,zhou2007,Smith2009,funk2010,morrison2016,Obertas2017} and typically use Equal Mutual Separation (hereafter EMS) as the system architectures, where the semi-major axis of adjacent planet pairs is determined by
\begin{equation}
a_{i+1}-a_i=K\Delta_H.
\end{equation}
and
\begin{equation}
\Delta_H=\frac{(a_{i}+a_{i+1})}{2}(\frac{m_{i}+m_{i+1}}{3M_\star})^{1/3}.
\end{equation}
where $\Delta_H$ is the mutual Hill radius, $a_i$ is the semi-major axis of the $i_{\rm th}$ planet, $m$ is the planetary mass, and $K$ is a numerical spacing parameter.  In this way, the period ratio between adjacent planets is:
\begin{equation}
\frac{P_{i+1}}{P_i} \approx 1+\frac{3}{2}K(\frac{m_{i}+m_{i+1}}{3M_\star})^{1/3}.
\end{equation}

For EMS planetary systems, $K$ is constant within one system.  This quantity is a key factor in determining the stability timescale $\tau$ of EMS planetary systems, where $\rm log\tau$ $\propto$ $K$ (see \citet{pu2015} for a review).  Another measure of the compactness of a system is the orbital period ratios between the planets.  Even with differences in planetary masses, it is clear that planet pairs with smaller period ratios are more compact and can be more strongly perturbed throughout their dynamical history than those with larger period ratios.  We see in Figure \ref{observation} that there is an obvious decrease of planet pairs towards small period ratios ($<1.5$), which may be caused, at least in part, by dynamical instability---a conjecture we investigate here. 

\citet{Izidoro:2017} studied the influence of dynamical instability on the period ratio distribution of multi-planet systems starting from compact resonant chains. But in this paper we conduct numerical simulations on non-EMS planetary systems with uniformly-distributed initial period ratios which we then evolve to determine the role that instability plays in shaping the final period ratio distribution.  By comparing the final distribution to the observed distribution, we should gain insight not only into the effects of dynamical instability, but also into the planet formation process generally.  That is, at least a portion of the difference between our simulations and the observations must be a consequence of the formation process itself, independent of the system's subsequent dynamical evolution. 

We describe our simulation techniques and the initial conditions in section \ref{sec:set-up}.  In section \ref{sec:stability}, we analyze the factors that influence the dynamical stability of multi-planet systems.  The consequences that instability has on the period ratio distribution and a comparison of the probability density function between our simulations and the de-biased \kepler\ observations are presented in section \ref{sec:periodratio}.  Finally, our conclusions are outlined in section \ref{sec:conclusion}.

\section{Simulation setup}\label{sec:set-up}

We consider four kinds of planetary systems containing $N$ planets orbiting a one solar mass star, where $N$ ranges from two to five.  We begin with samples of 1000 realizations for each kind of system.  The period ratio for each adjacent pair is assigned such that the period ratios in each suite are strictly uniform between one and three.  For example, for the five-planet systems, there are 4000 period ratios.  We therefore generate an array with 4000 equally-spaced elements between one and three and randomly choose the period ratios in each system from that array (without replacement) until the sample of 1000 systems is complete (as opposed to drawing period ratios from a uniform distribution, which would be subject to unwanted statistical variation).  Thus, our planetary systems are non-EMS and the distribution of period ratios for all adjacent planet pairs are uniform.  For each system, the innermost planet has an orbital period of 10 days, consistent with the typical orbital period of planets observed by the \kepler\ mission \citep{Thompson:2018}.  The eccentricity and inclination (in radians) for each planet are drawn from a Rayleigh distribution with $\sigma=10^{-3}$:
\begin{equation}
P(x)=\frac{x}{\sigma^2} e^{-\frac{x^2}{2\sigma^2}}.
\end{equation}
The planetary masses are also Rayleigh distributed with $\sigma_m=$ 6 $m_{\oplus}$,  based on the TTV mass of the \kepler\ observations\citep{Hadden2017}.  The minimal planetary mass is limited to be 1 $m_{\oplus}$.  Other orbital elements are randomly distributed between 0 and $360^{\circ}$.

We integrate each system up to $10^6$ years using the  $ias15$ integration scheme of the REBOUND package \citep{Rein2012,Rein2015}.  This integration time is about $3.65\times10^7$ orbits of the inner-most planet, $t_0$.  We include collisions in our integrations.  Once the distance between two planets is smaller than the sum of their planetary radius, they merge with momentum and mass conserved.  The planetary radius is calculated as $R=(m/(3m_{\oplus}))R_{\oplus}$ \citep{wu2013}.

\section{The stability criteria in multi-planet systems}\label{sec:stability}

\begin{figure}
\vspace{0cm}\hspace{0cm}
\centering
\includegraphics[width=\columnwidth]{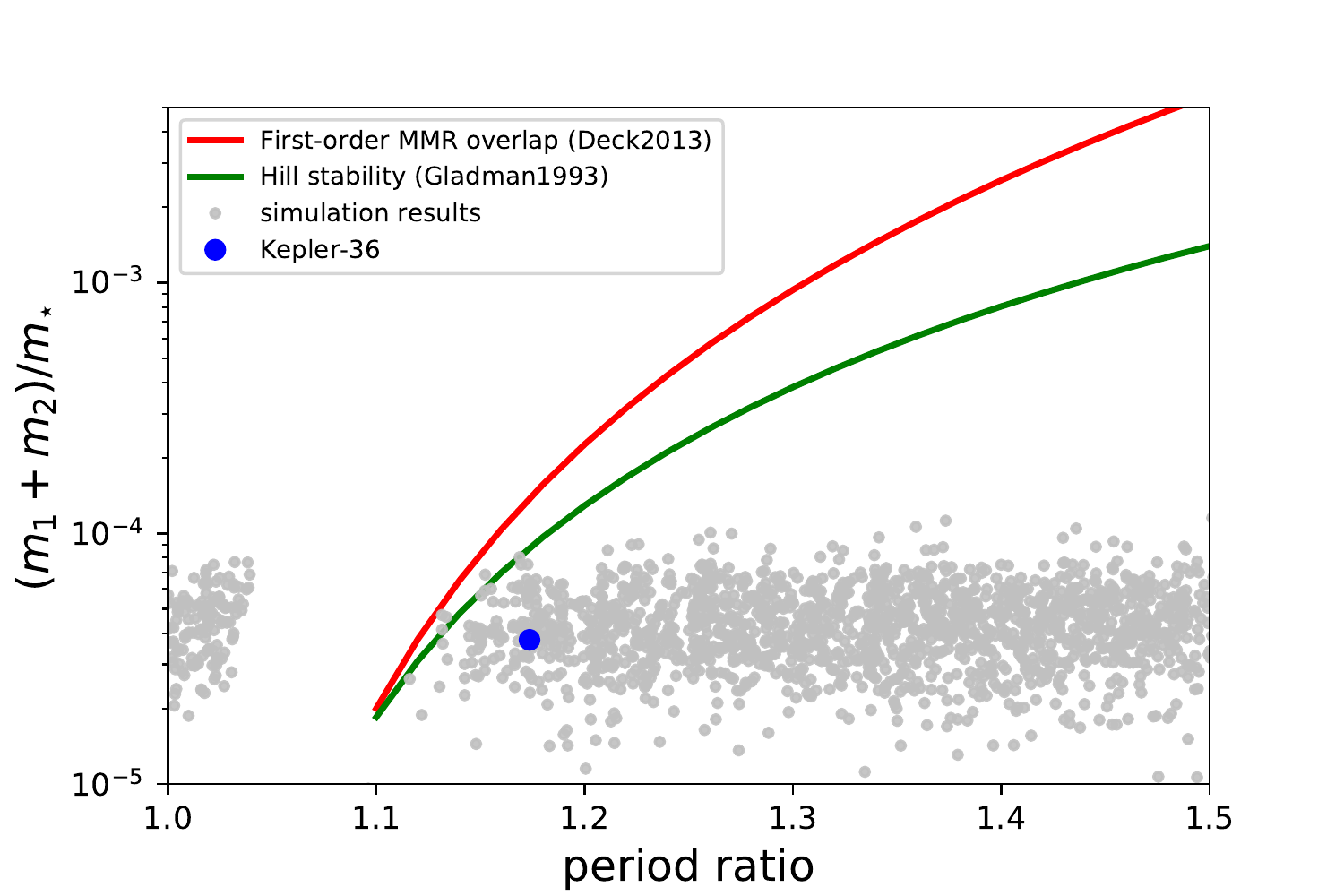}
\caption{The total planetary mass and period ratios of all adjacent planet pairs (shown as gray dots) in our simulation after $3.65\times10^7$ $t_0$.  The red curve is the first-order MMR overlap criteria of the initially circular case from \citet{deck2013}, the green curve is the Hill stability criteria from \citet{gladman1993}.  The blue dot represents the Kepler-36 system \citet{carter2012}.}
\label{pr_mass}
\end{figure}

The stability of planetary systems containing more than two planets is more challenging and less well understood than two planet systems.  Here, we study the stability criteria for both two-planet systems and systems with more than two planets.  The relationship between period ratio and planetary mass ($m_1+m_2$) of planet pairs in all four kinds of planetary systems after $3.65\times10^7$ $t_0$ is shown in Figure \ref{pr_mass}.  We find that the stable planet pairs have period ratios either smaller than 1.05 or larger than 1.1.  The two groups of planet pairs remain stable via different mechanisms, which are discussed in the following sections.

For planet pairs with period ratios larger than 1.1, two different criteria are often invoked to determine their stability, either the resonance overlap criteria \citep{wisdom1980,deck2013} or the Hill stability criteria \citep{gladman1993}.  We find that both criteria are reasonable approximations to the stability cutoff, but that the resonance overlap criteria performs better (it is strictly obeyed in our simulations for period ratios larger than 1.1).  For period ratios smaller than 1.05, the systems are stable if they are in the 1:1 MMR.

\subsection{Planet pairs in the 1:1 MMR}\label{sec:11mmr}

After an integration time of $3.65\times10^7 \ t_0$, some planet pairs with period ratios near 1 remain because they are protected by the 1:1 MMR.  Co-orbital configurations have been studied extensively, especially in planet-satellite systems \citep{dermott1981,dermott19812,yoder1983,tabachnik2000,christou2011}.  These insights are also applied to the problem where a terrestrial planet co-orbits with a gas giant \citep{dvorak2004,erdi2005,beauge2007}.  More general problems such as two comparable planets in 1:1 resonance have also been studied \citep{nauenberg2002,laughlin2002}.  

\begin{figure}
\vspace{0cm}\hspace{0cm}
\centering
\includegraphics[width=\columnwidth]{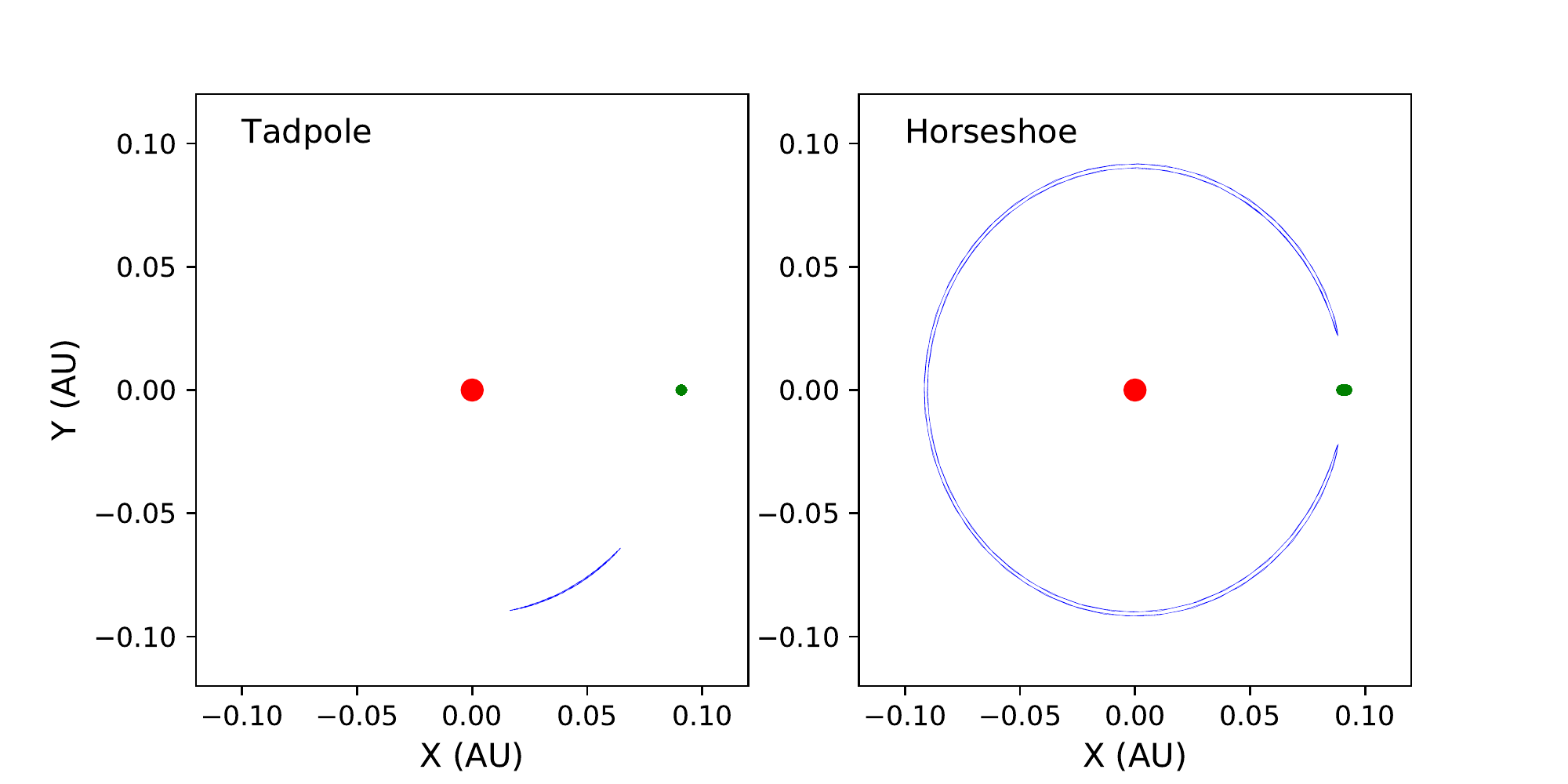}
\caption{ Position of one planet (shown as light blue) in the reference frame co-rotating with the other planet (shown as green dot).  The host star is shown as the red dot.  The left panel is the tadpole orbit and the right panel is the horseshoe orbit.}
\label{orbit}
\end{figure}

Of the stable, co-orbital planetary systems, planet pairs with initial differences of mean longitude far from $180^{\circ}$ and period ratios very close to 1 evolve in tadpole orbits (shown in the left panel of Figure \ref{orbit}), while planet pairs with period ratios slightly farther from 1 have horseshoe orbits (shown in the right panel of Figure \ref{orbit}). The fraction of tadpole orbits among all co-orbital configurations is about $25\%$.  (Recall that all of these co-orbital systems were generated randomly from our distributions of initial parameter values). 

Planet pairs in systems with more than two planets account for $87\%$ of all co-orbital configurations.  Hence, co-orbital planets are also likely to be stable in multi-planet ($N>2$) systems.   We check and find that planet pairs survived 1:1 MMR generally have period ratios $<1.03$. For planetary systems with more than two planets, the period ratio between the co-orbital pair and their closest companion should be larger than $1.33$ to ensure the stability of the co-orbital pair. The resonant angle $\phi=\lambda_2-\lambda_1$ of planet pairs in tadpole orbits (where $1$ and $2$ represent the two planets in the resonance) oscillates within a small range and one planet never crosses the $L_{3}$ Lagrange point of the other.  For planet pairs in horseshoe orbits, the resonant angles oscillate over a large range $>180^{\circ}$ of values---where one planet crosses the $L_3$, $L_4$ and $L_5$ Lagrange points of the other planet.

The co-orbital configuration of the two-planet case can be stable for as long as $3.65\times10^9$ $t_0$ (possibly longer), for both the tadpole and the horseshoe orbits.  \citet{tabachnik2000} showed that the Earth tadpole can be stable for as long as $10^9$ years, while horseshoe orbits are generally considered less stable than tadpole orbits \citep{dermott19812}.  \citet{laughlin2002} suggested that the horseshoe configuration can be stable for a long time if $(m_1+m_2)/m_{\star}\leq2\times10^{-4}$, which is the case for our simulations.  Although co-orbital planets were not found by \kepler\ \citep{janson2013}, \citet{ford2006} and \citet{leleu2017} proposed a method to detect them by combining transit and radial velocity measurements.  This method may have different detection sensitivities that may enable their discoveries in the future.  Nevertheless, if such planet pairs were common, they would likely have been detected by \kepler---especially in high Signal-to-Noise cases.  There are a few planet candidate systems that appear to have small period ratios such as KOI-284, KOI-521 and KOI-2248.  However, these systems show signs of being false positives, or (as in the case of KOI-284) false multis---where the signal is actually from two separate planetary systems in a stellar binary \citep{lissauer2014}.  Thus, we find it unlikely that co-orbital planet pairs are a common byproduct of planet formation.

\begin{figure}
\vspace{0cm}\hspace{0cm}
\centering
\includegraphics[width=\columnwidth]{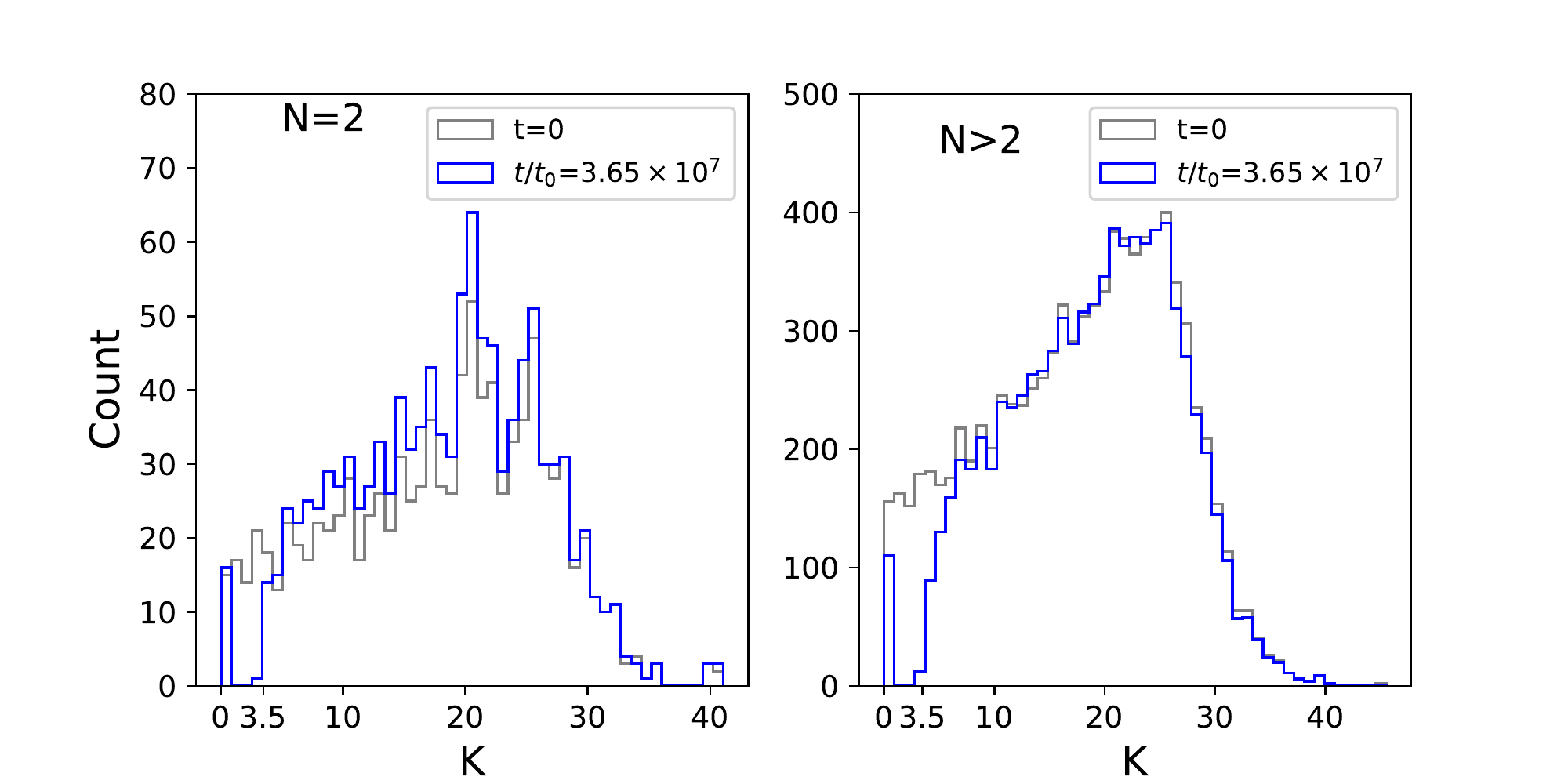}
\caption{The distribution of $K$ between each planet pair in the two-planet system (left panel) and systems with more than two planets (right panel).  The initial distributions are shown in gray and the final distributions are shown in blue.  Note that there are more two-planet systems after $t=3.65\times10^7$ $t_0$ because some of them are produced by planetary systems with more than two planets where collision or ejection occurs.}
\label{kdis}
\end{figure}

\subsection{Stability of planet pairs with period ratio $>$ 1.1}

We now turn from planets in the 1:1 MMR to pairs in multi-planet systems that have larger period ratios.  Previous works \citep{chambers1996,zhou2007,Smith2009,funk2010,pu2015,morrison2016,Obertas2017} have shown that the mutual separation in units of mutual hill radius, $K$, is one indicator of the instability timescale of EMS planetary systems.  \citet{gladman1993} showed that for two planet systems, the minimal $K$ required to remain stable is $\sim 3.5$.  Figure \ref{kdis} shows the initial and final $K$ distributions for systems with two or more planets.  The initial values of $K$ are distributed between 0 and 40.  After $3.65\times10^7 \ t_0$, however, $K$ of the remaining pairs are either very close to 0 or larger than the predicted stability cutoff of 3.5. For EMS planetary systems,  numerical simulation results in \citet{Obertas2017} show that five planet systems can survive at least $10^9$ $t_0$ for $K\geq8.5$. The criterion $K>3.5$ between each planet pair alone can not ensure the stability of the multi-planet (N$>$2) systems. For non-EMS planetary systems, we investigate whether $K$ between each planet pair, or some other statistic derived from $K$, best characterizes the stability of the multi-planet systems in the following sections.

\subsubsection{Factors that determine the stability in multiple planet systems}

\begin{figure}
\vspace{0cm}\hspace{0cm}
\centering
\includegraphics[width=\columnwidth]{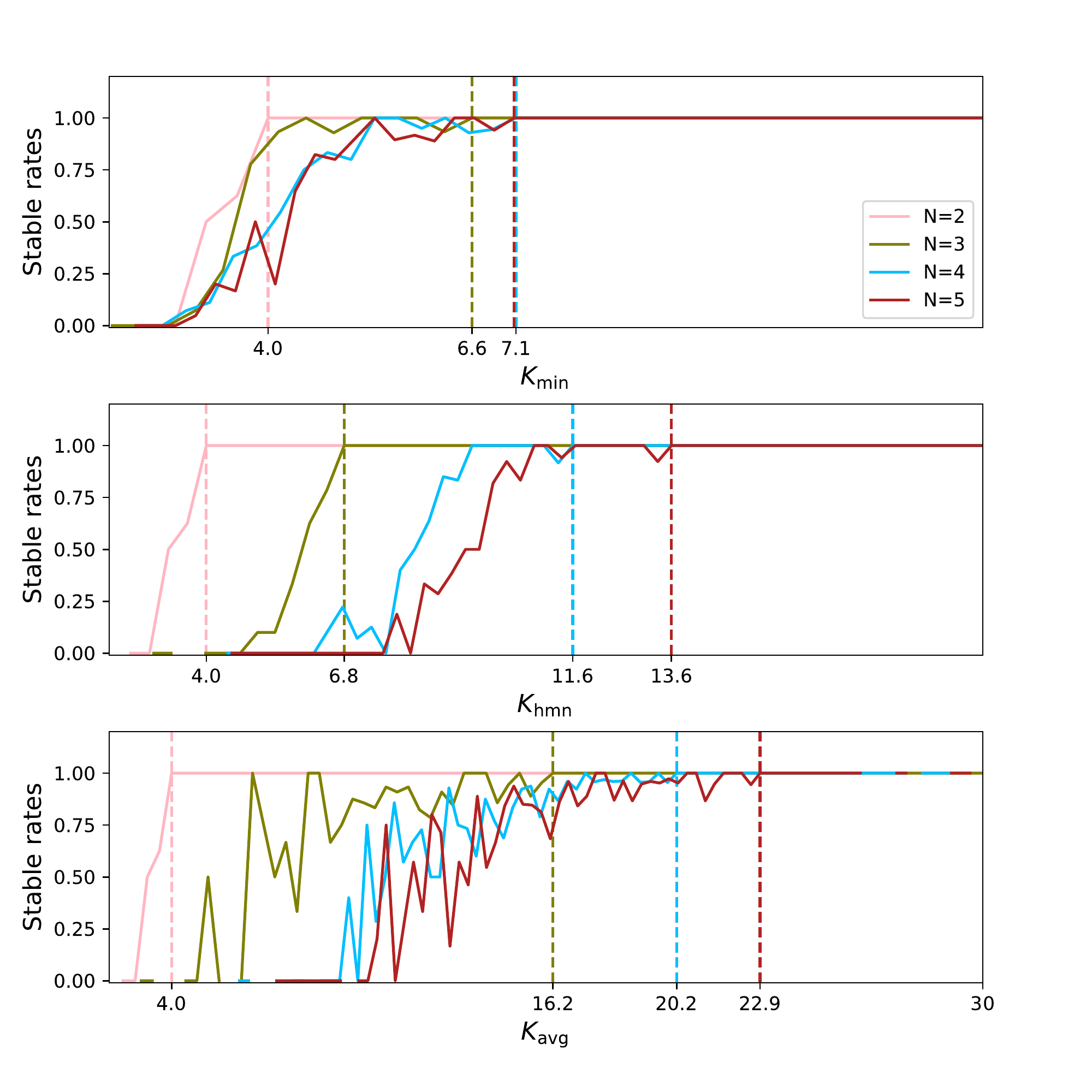}
\caption{The upper, middle and lower panels show stable rates as a function of $K_{\rm min}$, $K_{\rm hmn}$ and $K_{\rm avg}$, respectively.  Different colors represent different kind of planetary systems.  Two planet systems are shown in light red, three planet systems are shown in yellow, four planet systems are shown in light blue and five planet systems are shown in dark red.  Each kind of planet pair is divided into 100 groups based on $K_{\rm min}$, $K_{\rm hmn}$ and $K_{\rm avg}$, respectively, and the stable rates are calculated in these small groups. }
\label{stable_rates}
\end{figure}

We consider three statistics derived from $K$: the minimum $K$ in a system ($K_{\rm min}$),  the harmonic mean value of $K$ ($K_{\rm hmn}$), and the arithmetic mean value of $K$ ($K_{\rm avg}$).  
Generally, the minimum mutual separation ($K_{\rm min}$) represents the local compactness of the planetary system, with the other two means gradually transitioning between local compactness and global compactness (the harmonic mean is the smallest of the three Pythagorean means and the arithmetic mean is the largest).  The stable rates of planetary systems at different $K_{\rm min}$, $K_{\rm hmn}$, and $K_{\rm avg}$ are shown in Figure \ref{stable_rates}.  Here, our measure of the stability of a planetary system is whether or not the planetary orbits remain near their initial values throughout the integration.  That is, $\left|P_{f}-P_{i}\right|<0.01P_{i}$, where $P_{i}$ and $P_{f}$ represent the initial and final orbital period, respectively.  

We see that the stable rates increase with all three statistics $K_{\rm min}$, $K_{\rm hmn}$, and $K_{\rm avg}$.  Once $K_{\rm min}$, $K_{\rm hmn}$ or $K_{\rm avg}$ exceeds a particular critical value (noted as $K_{\rm min, crit}$, $K_{\rm hmn, crit}$ and $K_{\rm avg, crit}$, respectively), the stability rates are $100\%$, meaning that the planetary system is stable for at least $3.65\times10^7 \ t_0$.  The critical values for the three statistics of $K$ are shown in Table \ref{tab1}.  For the two-planet systems, the critical values of the $K$'s are all near 4 with uncertainties of 0.4---slightly larger than the traditional 3.5.  Part of the reason for this larger cutoff may be that we have very few samples of planetary systems around 3.5, and our stability criteria is quite restrictive.  Increasing the number of planets within one system increases the critical values of $K_{\rm min}$, $K_{\rm hmn}$, and $K_{\rm avg}$.

Among all stable planetary systems, the fraction of planetary systems with $K_{\rm min}>K_{\rm min, crit}$, $K_{\rm hmn}>K_{\rm hmn, crit}$ and $K_{\rm avg}>K_{\rm avg, crit}$ are shown in Table \ref{tab1}, respectively.  A large fraction of stable systems above the critical value indicates a good stability criterion since it places a better constraint on the stable spacings of planets.  We see from Table \ref{tab1} that $K_{\rm avg}$ is not a good statistic to determine the stability of multi-planet systems, especially for systems containing four or more planets---nearly 85\% of stable systems have separations smaller than the threshold where all systems are seen to be stable.    
For $K_{\rm min}$ and $K_{\rm hmn}$, only about 25\% of the stable systems are below the threshold.  To better determine which of these statistics best constrains the dynamics of the system, we move on to compare the instability timescales determined by $K_{\rm min}$ and $K_{\rm hmn}$ between our samples and the EMS systems.

\begin{table}
\begin{center}
\caption{Critical values of different statistics of $K$ for planetary systems containing different number of planets (the second, third and fourth row) and fraction of planetary systems meeting the criteria among all stable planetary systems (the fifth, sixth and seventh row).\label{tab1}}
\resizebox{\linewidth}{!}{%
\begin{tabular}{ccccc}
\hline
\hline
&N=2&N=3&N=4&N=5\\\hline
$K_{\rm min, crit}$&$4.0\pm0.4$&$6.6\pm0.3$&$7.1\pm0.3$&$7.1\pm0.3$\\\hline
$K_{\rm hmn, crit}$&$4.0\pm0.4$&$6.8\pm0.4$&$11.6\pm0.3$&$13.6\pm0.3$\\\hline
$K_{\rm avg, crit}$&$4.0\pm0.4$&$16.2\pm0.4$&$20.2\pm0.3$&$22.9\pm0.3$\\\hline
$K_{\rm min}>K_{\rm min, crit}$&$99.2\%\pm0.5\%$&$88.2\%\pm1.0\%$&$80.7\%\pm2.0\%$&$75.8\%\pm2.7\%$\\\hline
$K_{\rm hmn}>K_{\rm hmn,crit}$&$99.2\%\pm0.5\%$&$97.9\%\pm1.3\%$&$82.9\%\pm2.0\%$&$72.5\%\pm2.4\%$\\\hline
$K_{\rm avg}>K_{\rm avg,crit}$&$99.2\%\pm0.5\%$&$69.1\%\pm2.5\%$&$38.9\%\pm2.4\%$&$15.9\%\pm1.5\%$\\\hline
\end{tabular}}
\end{center}
\end{table}

\subsubsection{Lower limit of instability timescale determined by $K_{\rm min}$}

\begin{figure}
\vspace{0cm}\hspace{0cm}
\centering
\includegraphics[width=\columnwidth]{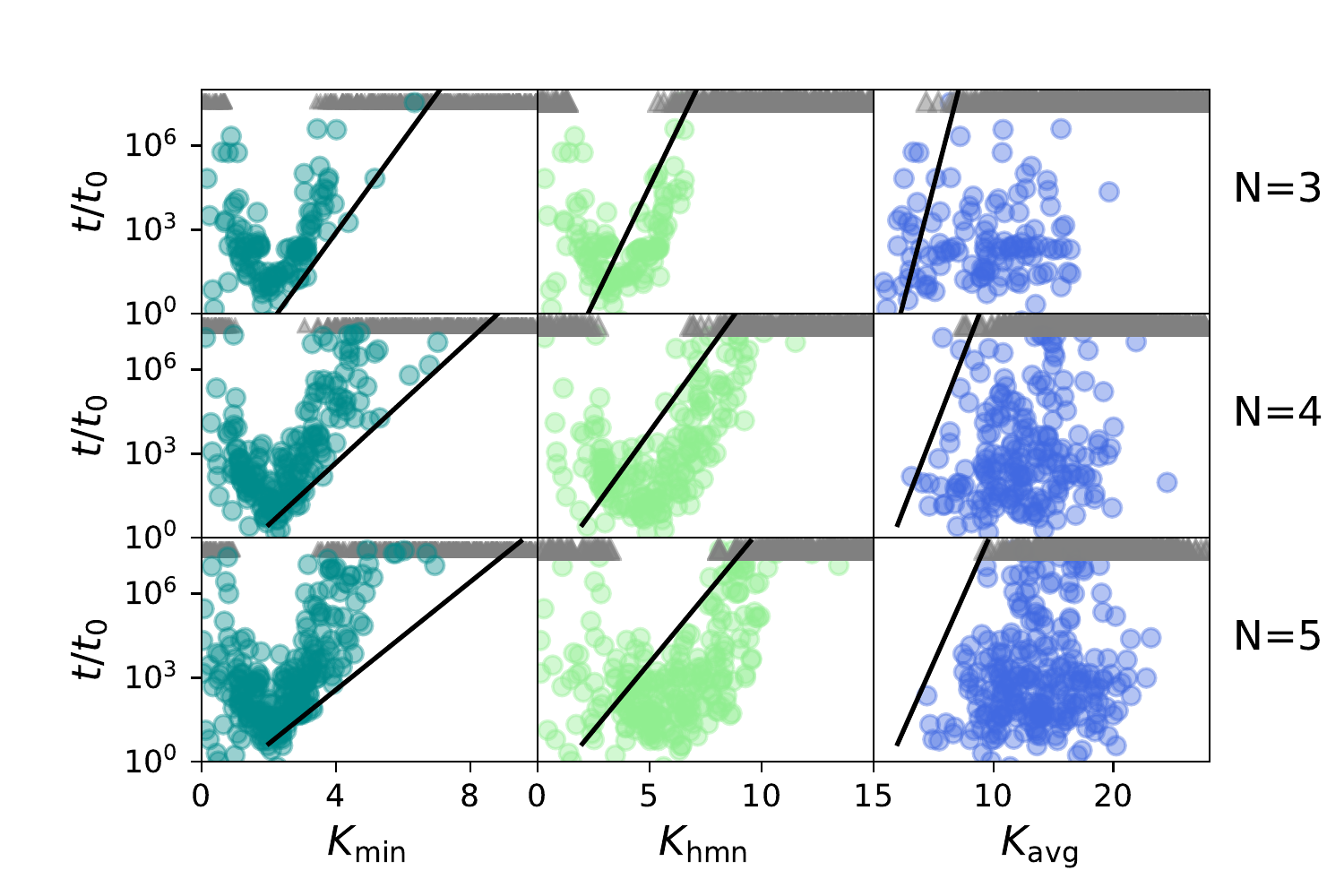}
\caption{The instability timescale $t/t_0$ vs.  $K_{\rm min}$ (left panel), $K_{\rm hmn}$ (middle panel) and $K_{\rm avg}$ (right panel).  The upper, middle and lower panel shows the results of three, four and five planet systems, respectively.  Systems that have close encounters are shown as dots, while systems that are stable within $3.65\times10^7$ $t_0$ are shown as triangles.  Dark lines show the instability timescale (from previous works) as a function of $K$ for EMS planetary systems.  For three planet systems $log_{10}t/t_0=1.65K-3.71$ \citep{chambers1996}, for four planet systems $log_{10}t/t_0=1.10K-1.75$ \citep{Rice2018}, for five planet systems $log_{10}t/t_0=0.964K-1.289$\citep{Obertas2017}.  }
\label{tk1}
\end{figure}

In this section, we calculate the instability timescale when a first close encounter occurs in our simulations.  Figure \ref{tk1} compares our results to the results from EMS systems in \citet{chambers1996}, \citet{Obertas2017} and \citet{Rice2018}.  We can see that the instability timescales for these systems have a large scatter, even at the same $K_{\rm min}$, $K_{\rm hmn}$, or $K_{\rm avg}$.  However, the lower limit of the instability timescale at different $K_{\rm min}$ is consistent with the value calculated in EMS systems.  At $K_{\rm min}>2$, we can determine a lower bound on the stability timescale for the system.  When using $K_{\rm hmn}$ and $K_{\rm avg}$, the estimated instability timescale no longer yields a good lower bound on the measured timescale, especially for $10<K_{\rm avg}<20$ where the instability timescale varies between 1 and $3.65\times10^7 \ t_0$. However, we can estimate the upper bound of the instability timescale with $K_{\rm hmn}$ or $K_{\rm avg}$. A combination of $K_{\rm min}$ and $K_{\rm hmn}$ (or $K_{\rm avg}$) would yield the variation in instability timescale.  \citet{pu2015} also conduct numerical simulations on non-EMS systems, they drew the value of K from a Gaussian distribution with mean value $K_{\rm mean}$ and variance $\sigma_K$, and they found that the instability timescale is well determined by $K_{\rm mean}-0.5\sigma_K$. Since $K_{\rm mean}-0.5\sigma_K$ is close to the smaller values of K in each system, our results are consistent.

\begin{figure}
\vspace{0cm}\hspace{0cm}
\centering
\includegraphics[width=\columnwidth]{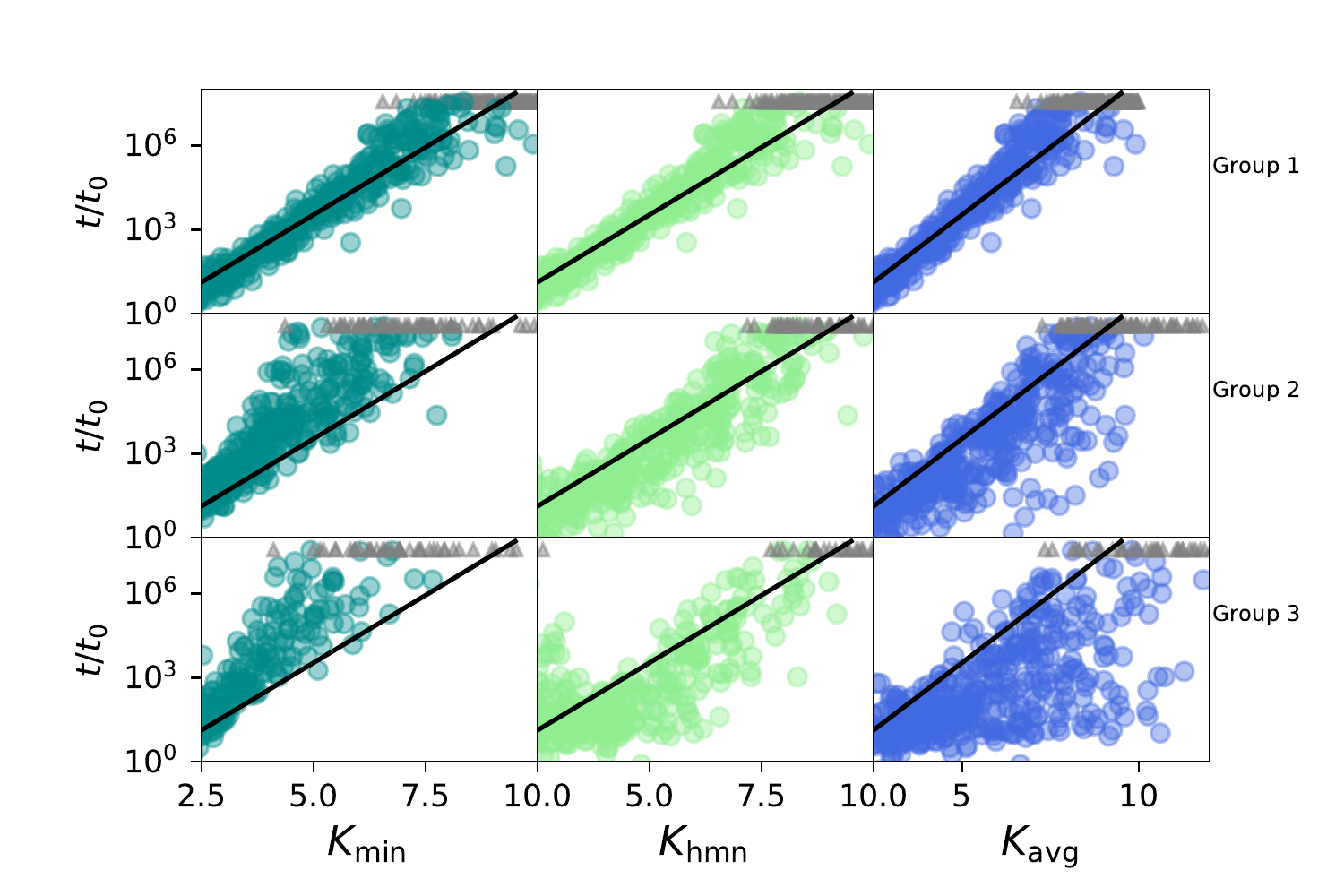}
\caption{The instability timescale $t/t_0$ vs.  $K_{\rm min}$ (left panel), $K_{\rm hmn}$ (middle panel) and $K_{\rm avg}$ (right panel) for the five-planet systems.  Systems that have close encounters are shown as dots, while systems that are stable within $3.65\times10^7$ $t_0$ are shown as triangles.  Upper panels: Group 1, the EMS systems.   Middle panels: Group 2, median values of $K$ ($K_m$) in each planetary system are uniformly distributed between 2.5 and 10, the standard deviation of $K$ in each system is $\sigma_K=0.3$ $K_m$. Lower panels: Group 3, similar to middle panels, but with $\sigma_K=0.6$ $K_m$.  The dark lines represent the instability timescale as a function of $K$, i.e.,$log_{10}t/t_0=0.964K-1.289$ from \citet{Obertas2017}.}
\label{tk2}
\end{figure}

As the $K$'s in our simulations are mostly distributed between 10  and 30 (as shown in Figure \ref{kdis}), with only a few examples of $K$'s between 2.5-10, we carry out a set of additional simulations focusing on small separations. 
We simulate three groups of five-planet systems using the same distribution of orbital elements and planetary mass as those described in Section \ref{sec:set-up}, except now we change their distribution of orbital periods.  We consider three different scenarios.  Group 1: we adopt EMS systems where the $K$'s are uniformly distributed between 2.5 and 10. Group 2: the median values of $K$ in each planetary system ($K_m$) are uniformly distributed between 2.5 and 10 and the standard deviation of $K$ in each system is $\sigma_K=0.3 \ K_m$. Group 3: similar to Group 2, but with $\sigma_K=0.6 \ K_m$.  The instability timescales of the three groups are shown in Figure \ref{tk2}.

For Group 1, the EMS case, our results agree with those of \citet{Obertas2017}, although the planets in our simulations have different masses.  For Groups 2 and 3, the scatter in the instability timescale is as large as four orders of magnitude, much larger than what is observed with the equal spacing of Group 1. Also, the scatter in instability timescale increases with the scatter of $K$ in each planetary system.  Despite the large scatter, we can still approximate the lower limit of the instability timescale with $K_{\rm min}$ using the relationship between $log_{10}t/t_0$ and $K$ in EMS systems.  The variation of instability timescale of one planetary system can be roughly determined with $K_{\rm min}$ and $K_{\rm hmn}$ (or $K_{\rm avg}$).  Nevertheless, with the ability of determining the lower limit of the instability timescale, $K_{\rm min}$ performs better than $K_{\rm hmn}$ and $K_{\rm avg}$ as a stability criteria in combination with the analysis in previous paragraphs.

\section{Period ratio distribution}\label{sec:periodratio}

\begin{figure}
\vspace{0cm}\hspace{0cm}
\centering
\includegraphics[width=\columnwidth]{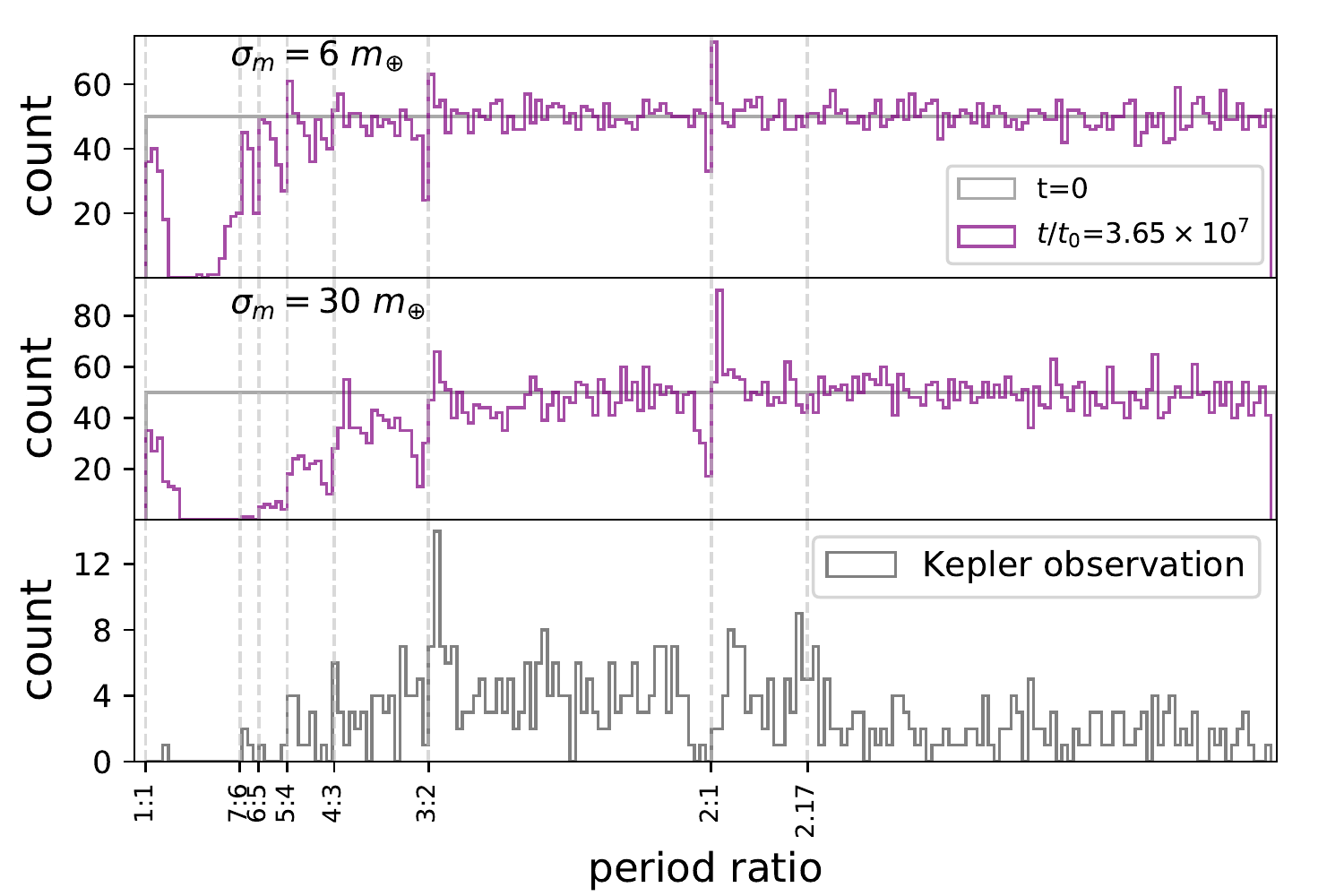} 
\caption{ The upper panel is the initial (shown in gray) and final (shown in purple) period ratio distribution from our simulations described in Section \ref{sec:set-up} where the planetary mass are Rayleigh distributed with $\sigma_m=6$ $m_{\oplus}$.  The middle panel is the initial (gray) and final (purple) period ratio distribution from the simulations where the planetary mass are Rayleigh distributed with $\sigma_m=30$ $m_{\oplus}$. The lower panel is the period ratio distribution of the confirmed \kepler\ planet pairs.  The vertical dashed lines indicate planet pairs near the  first order MMRs and period ratio of 2.17.  $t_0$ is the initial orbital period of the inner most planet.  }
\label{pr_Rl}
\end{figure}

Planet pairs with small $K_{\rm min}$ likely collide with each other or are scattered, and as a consequence, the architecture of multi-planet systems are sculpted by their dynamical evolution.  Here, we study the final period ratio distribution of the systems after $3.65\times10^7 \ t_0$.  The initial and final period ratio distributions of all planetary systems are shown in the upper panel of Figure \ref{pr_Rl}.  

We can see in that figure that planet pairs with period ratios smaller than 1.05 or larger than 1.1 remain stable.  Planet pairs with period ratios near 1 are protected by the 1:1 MMR, as discussed in Section \ref{sec:11mmr}.  The number of stable planetary systems increases with period ratio between 1.1 up to a value near 1.33, after which the distribution is almost flat.  Additionally, we see that there are dips on the near side and peaks on the far side of the first-order MMRs, including 2:1, 3:2, 4:3, 5:4, 6:5, and 7:6.  This result is similar to the observed period ratio distribution (lower panel of Figure \ref{pr_Rl}), except that the width and depth of the gap on the near side of the MMRs are smaller than those in the observation. Also, there is no significant feature at period ratio of 2.17 in the simulation. Period ratio distribution from Pu $\&$ Wu (2015) also shows asymmetry features around MMRs, but there is no obvious peaks on the far side of MMRs in their simulations. We investigate how these features were produced in the following paragraphs.

\subsection{Period ratio asymmetry near first-order MMR}

The behavior of two accreting planets near the first order MMRs 2:1 and 3:2 has been studied by \citet{petrovich2013}.  They found that the period ratio distribution develops an asymmetric dip-peak structure near the resonance. In our simulations, this feature appears in both two planet systems and systems with more than two planets, although the planetary mass is fixed during the evolution.  We find that planet pairs with initial period ratios near MMR are likely to have final period ratios larger than their initial values.

\begin{figure}
\vspace{0cm}\hspace{0cm}
\centering
\includegraphics[width=\columnwidth]{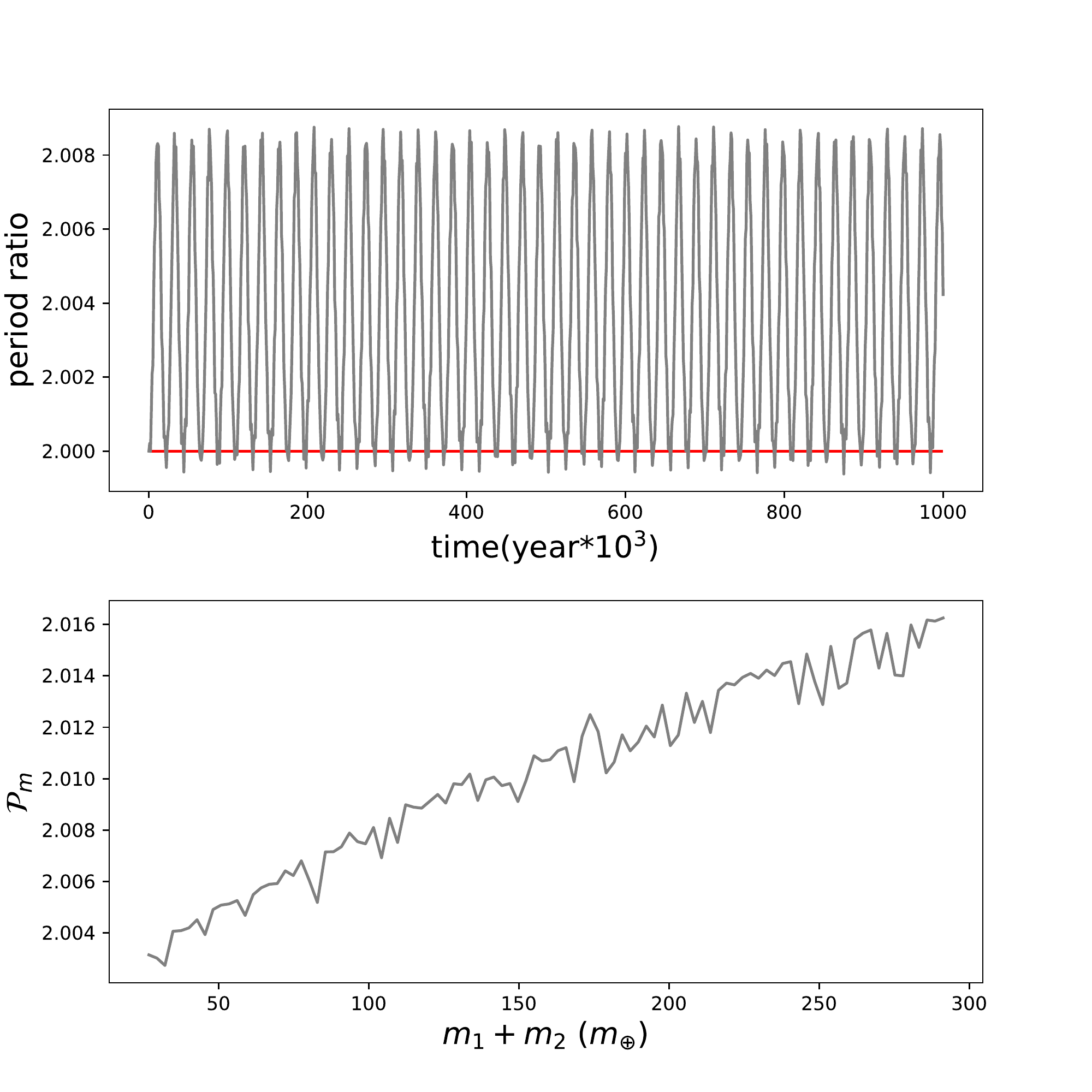}
\caption{ Upper panel: The evolution of the period ratio (shown as the gray curve) with initial value of 2.0 (shown as the red horizontal line) in one of the two-planet systems.  Lower panel: the median value of the period ratio $\mathcal{P}_m$ during $3.65\times10^7 \ t_0$ with different total planetary mass $m_1+m_2$.  The total mass is randomly split between $m_1$ and $m_2$. Other orbital elements are the same with the example in the upper panel.}
\label{pr_evo}
\end{figure}

\begin{figure}
\vspace{0cm}\hspace{0cm}
\centering
\includegraphics[width=\columnwidth]{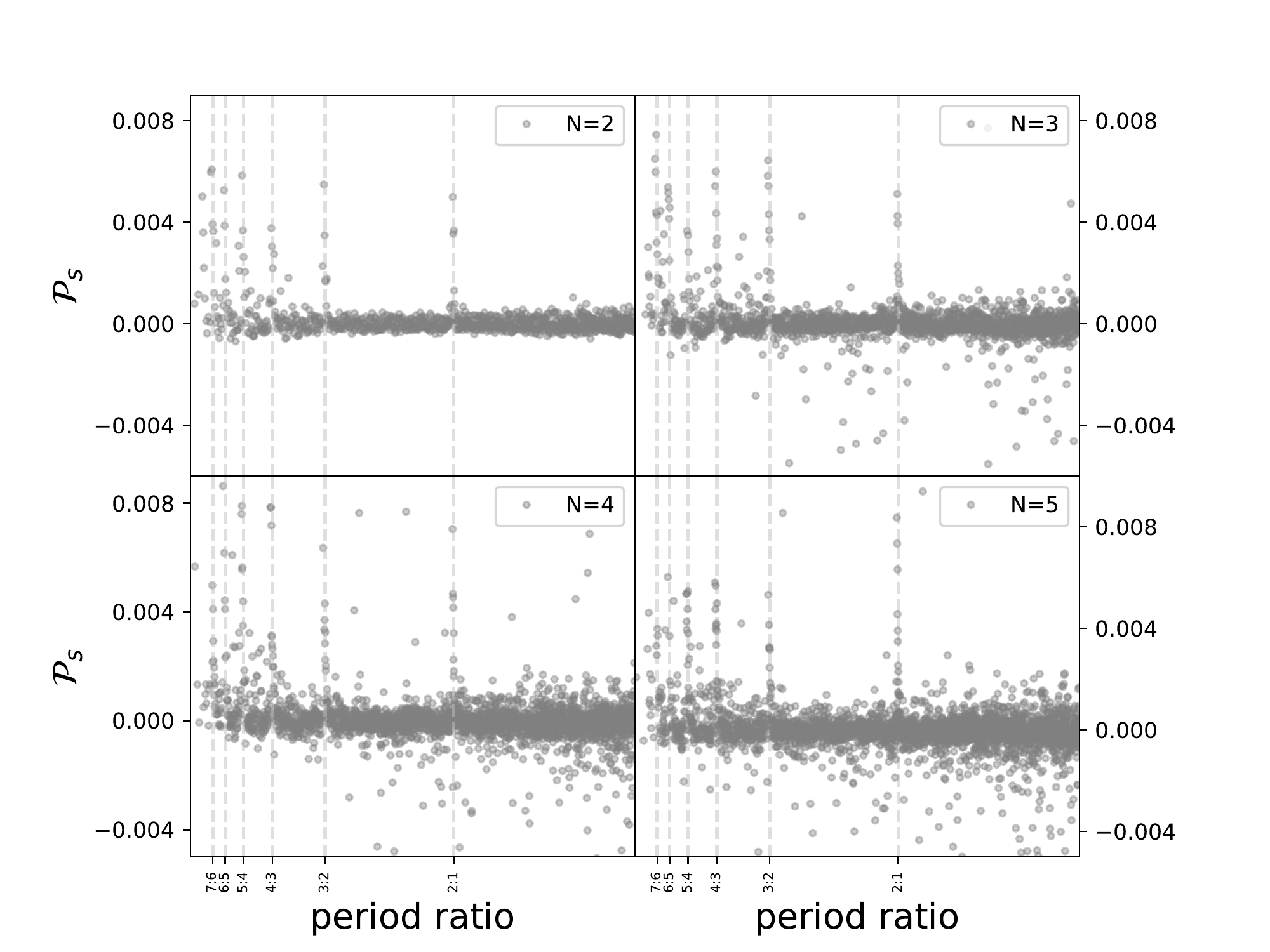}
\caption{The average difference $\mathcal{P}_s$ between the period ratio during the evolution and the initial period ratio at different initial period ratios in different planetary systems.  The vertical dashed lines indicate the position of first order MMRs.  $\mathcal{P}_s>0$ indicates that the planet pair is more likely to be on the far side of the initial period ratio than the near side.  }
\label{fig8}
\end{figure}

One example of a planet pair in the two-planet system with an initial period ratio of 2.0 is shown in the upper panel of Figure \ref{pr_evo}.  As its period ratio evolves due to mutual interaction, the pair tends to stay on the far side of the 2:1 MMR.  To better describe this property, we define the average difference between the period ratio during the evolution and the initial period ratio as $\mathcal{P}_s=\sum^{n}_{i=1}(pr_i-pr_0)/n$, where $n$ represents the number of data that is output during the simulation, $pr_i$ represents the period ratio of the $i_{\rm th}$ output from the simulation, and $pr_0$ represents the initial period ratio.  If $\mathcal{P}_s>0$, then the period ratio is more likely to be larger than its initial value.  We show $\mathcal{P}_s$ at different period ratios for two, three, four, and five planet systems in Figure \ref{fig8}.  We find that there are significant peaks of $\mathcal{P}_s$ at period ratios 7:6, 6:5, 5:4, 4:3, 3:2, and 2:1---especially for the two planet systems.  A consequence of this feature is that whenever we measure the period ratio distribution, there is excess probability that period ratios initially on the near side of the MMRs will be seen on the far side.

\citet{petrovich2013} proposed that the equivalent width of the peaks/dips is proportional to the planetary mass.  To verify this conclusion, we choose the same two-planet system shown in the upper panel of Figure \ref{pr_evo} to conduct an additional set of simulations. We slowly increase the total mass of the two planets and calculate the median value of the period ratio $\mathcal{P}_m$ during the evolution.  The total mass is randomly split between the two planets. We find that $\mathcal{P}_m$ does increase with the planetary mass, in agreement with their work (see the lower panel of Figure \ref{pr_evo}).

They also suggest that planetary mass should be in the range of $20-100$ $m_{\oplus}$ in order to explain the structure near 3:2 and 2:1 MMRs in the \kepler\ observation.  Such masses are much larger than the masses we use in our simulations and are larger than the planetary mass obtained for the typical \kepler\ system as measured with transit time variations \citep{Hadden2017}.  To verify that this dip-peak structure persists over a longer evolution time, we integrate the five-planet systems up to $3.65\times10^8 \ t_0$ for a comparison.  We found that the two results are similar as all of the features remain (except for an additional 18 systems that go unstable).

\subsubsection{Varying the planetary mass distribution}

As mentioned above, the planetary masses in our simulations are too small to fully explain the observations with this mechanism.  In this section, we used Rayleigh distributed planetary masses with $\sigma_m=30 \ m_{\oplus}$.  (The average value of planetary mass is increased by a factor of five from the previous section.)  The other parameter distributions remain the same.  The final period ratio distribution for these simulations is shown in the middle panel of Figure \ref{pr_Rl}.  We see that the widths and depths of the dips near the first order MMRs are larger than those of the smaller planetary mass with $\sigma_m=6 \ m_{\oplus}$---especially for the 2:1 and 3:2 MMRs where the gap on the near side is only slightly smaller than the observations.

Additionally, the increase of planetary mass by a factor of 5 leads to a decrease of $K$ by a factor of 1.7 from the original values, substantially reducing the instability timescale (particularly for planet pairs with period ratios between 1.1 and 1.5).  Since the planetary masses observed by \kepler\ are rarely this large, the mechanism we present here can only account for a portion of the observed asymmetry in the period ratio distribution near MMR.  In addition, we note that the shallower period ratio distribution for small period ratios could be used to constrain the planetary masses observed in \kepler\ systems---though the constraint from TTV observations is likely more stringent.



\subsubsection{Varying the eccentricity distribution}\label{ecc}

\begin{figure}
\vspace{0cm}\hspace{0cm}
\centering
\includegraphics[width=\columnwidth]{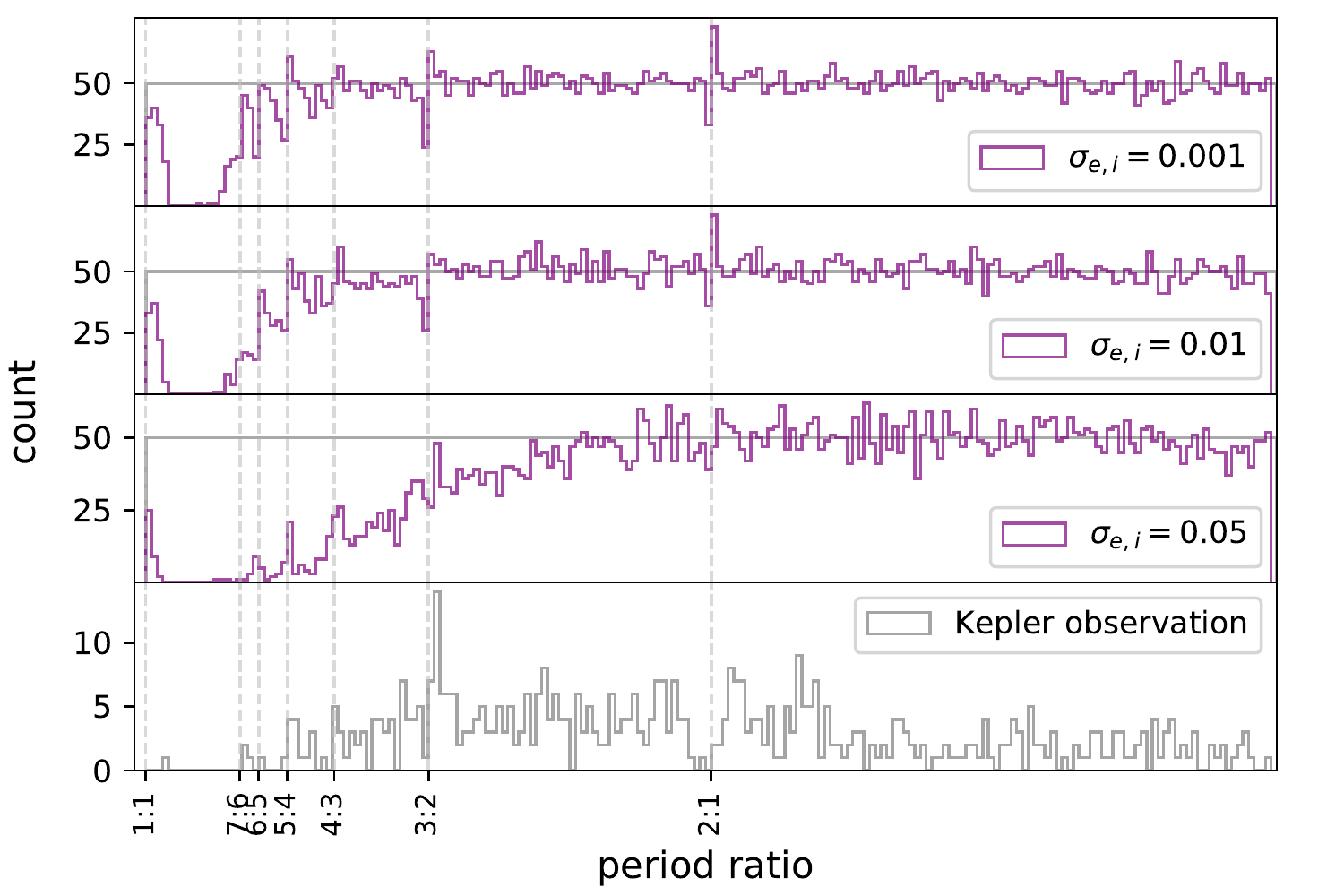}
\caption{The top three panels show the initial (light gray) and final (purple) period ratio distributions of our simulations with Rayleigh distributed eccentricity and inclination $\sigma_{e,i}=10^{-3}$, $\sigma_{e,i}=0.01$, $\sigma_{e,i}=0.05$, respectively. The bottom panel is the Kepler period ratio distribution. }
\label{pr_Rl_e}
\end{figure}

We now consider the effects of larger initial eccentricities and inclinations.  In previous sections, the orbits of the planets are nearly circular and co-planar with eccentricities and inclinations $\sim 10^{-3}$.  Here, we use eccentricity and inclination distributions of $\sigma_{e,i}=0.01$ and $\sigma_{e,i}=0.05$.  Again, other parameter distributions remain the same with those described in Section \ref{sec:set-up}.  The final period ratio distributions are shown in Figure \ref{pr_Rl_e}.  The results of the simulations with $\sigma_{e,i}=0.01$ are similar to those of $\sigma_{e,i}=1\times10^{-3}$, except that the peak on the far side of the 3:2 MMR is not as strong.  However, when $\sigma_{e,i}$ increases to 0.05, peaks and dips near MMRs almost disappear, as shown by \citet{Xie:2014} that the asymmetry features around MMRs will become weaker with increasing eccentricity.

Planets with higher eccentricities tend to be more unstable when their period ratios are between 1.1 and 1.7 than in the small eccentricity and inclination cases.  This is both because the increased eccentricity yields a higher probability that two planets have close encounters and because the resonance width increases with eccentricity \citep{deck2013,Hadden:2018}, so resonance overlap is more likely to occur.  We compare our simulations to the resonance overlap criteria from \citet{Hadden:2018} and find that our results conform to that stability criteria. Thus, distributions of eccentricity and inclination with $\sigma_{e,i}=0.05$ are too large for planet pairs to produce the observed features.  Moreover, the larger eccentricities and inclinations yield a period ratio distribution that is more shallow between 1.1 and 1.5 than the observations (similar to what occurred with larger mass planets from the previous section).  We investigate the constraints that can be placed on the eccentricities from this feature in a later section.


\subsection{The probability density function of the period ratios}\label{subsec:pdf}

\subsubsection{De-biased period ratios of the \kepler\ planets}

We have shown that (at least a portion of) the asymmetry feature near MMRs can be produced via planetary dynamics originating from a distribution that lacks those features.  In this section, we compare the probability density function (PDF) of period ratios between the observed \kepler\ data and our simulations.  As expected, \kepler\ observations contain geometric bias and pipeline incompleteness \citep{ragozzine2010,borucki2011,lissauer2011,ciardi2013,steffen2015,coughlin2016,brakensiek2016}.  \citet{steffen2015} suggests that the influence of pipeline incompleteness, compared to the geometric bias, is the smaller of the two effects so we only consider the geometric bias here.

We have a total of 583 confirmed planet pairs with period ratio $<5$ from the Q1-Q17 DR25 catalog.  To avoid the influence of very long period planets, which can significantly affect the distribution if not treated correctly, we cut off the sample with $a/R_{\star}<150$.  According to previous studies \citep{lissauer2011,fang2012,Tremaine2012,fabrycky2014}, \kepler\ multi-planet systems are rather flat, so we assume the mutual inclination of planets in a system are Rayleigh distributed with $\sigma\sim$ $1.5^{\circ}$, similar to \citet{steffen2015}.  We use the CORBITS algorithms from \citet{brakensiek2016} to calculate the probability of detecting the outer planet given that the inner planet is detected.  The inverse of the probability is adopted as the weight of the planet pair.  Finally, we construct a kernel density estimator of the period ratio distribution.  For each period ratio, we use a Gaussian distribution with the median value $\mu$ equal to the period ratio $Pr$ and the standard deviation $\sigma$ to be $0.00005Pr$.  The total area of the Gaussian distributions is normalized to 1.  

\begin{figure}
\vspace{0cm}\hspace{0cm}
\centering
\includegraphics[width=\columnwidth]{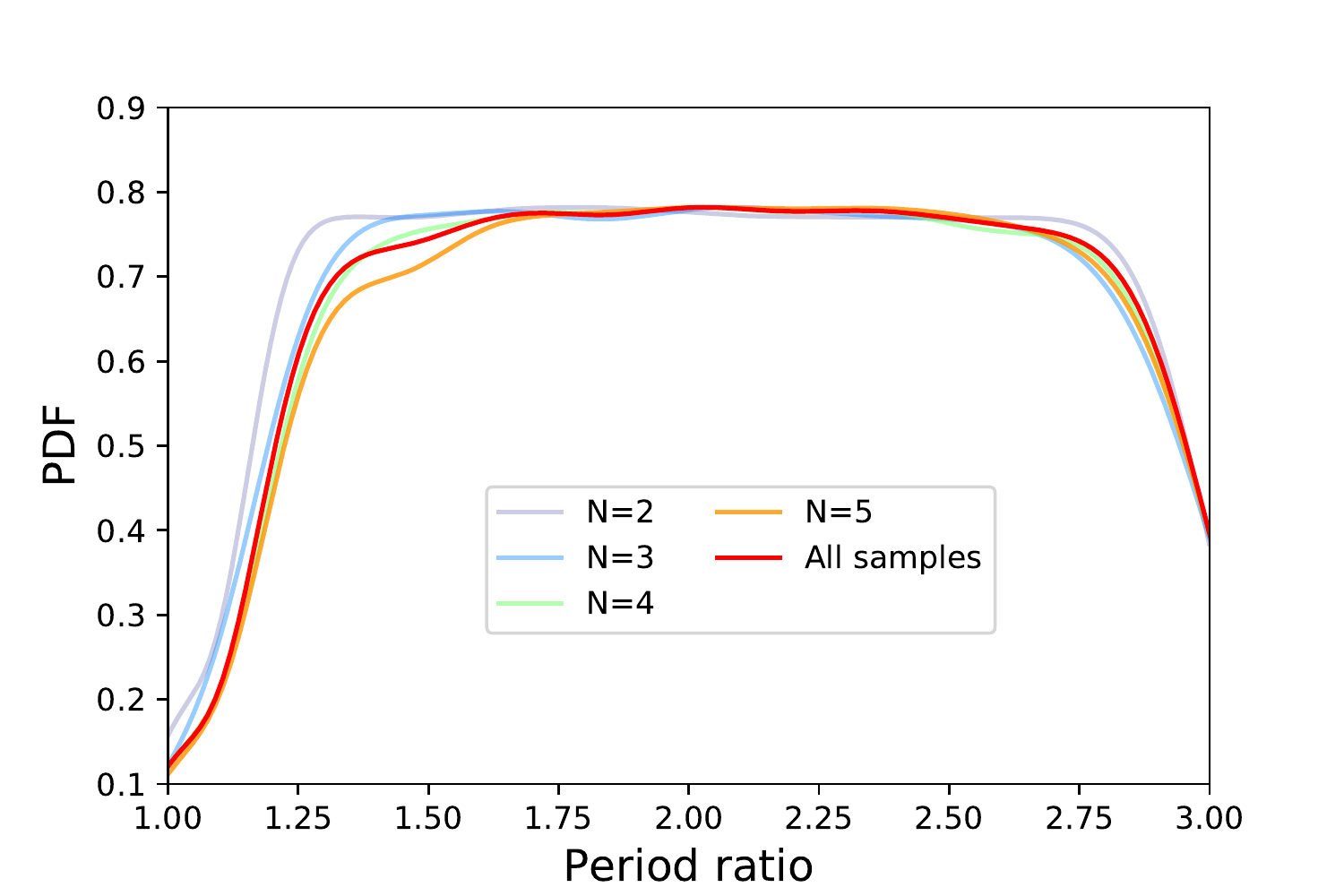}
\includegraphics[width=\columnwidth]{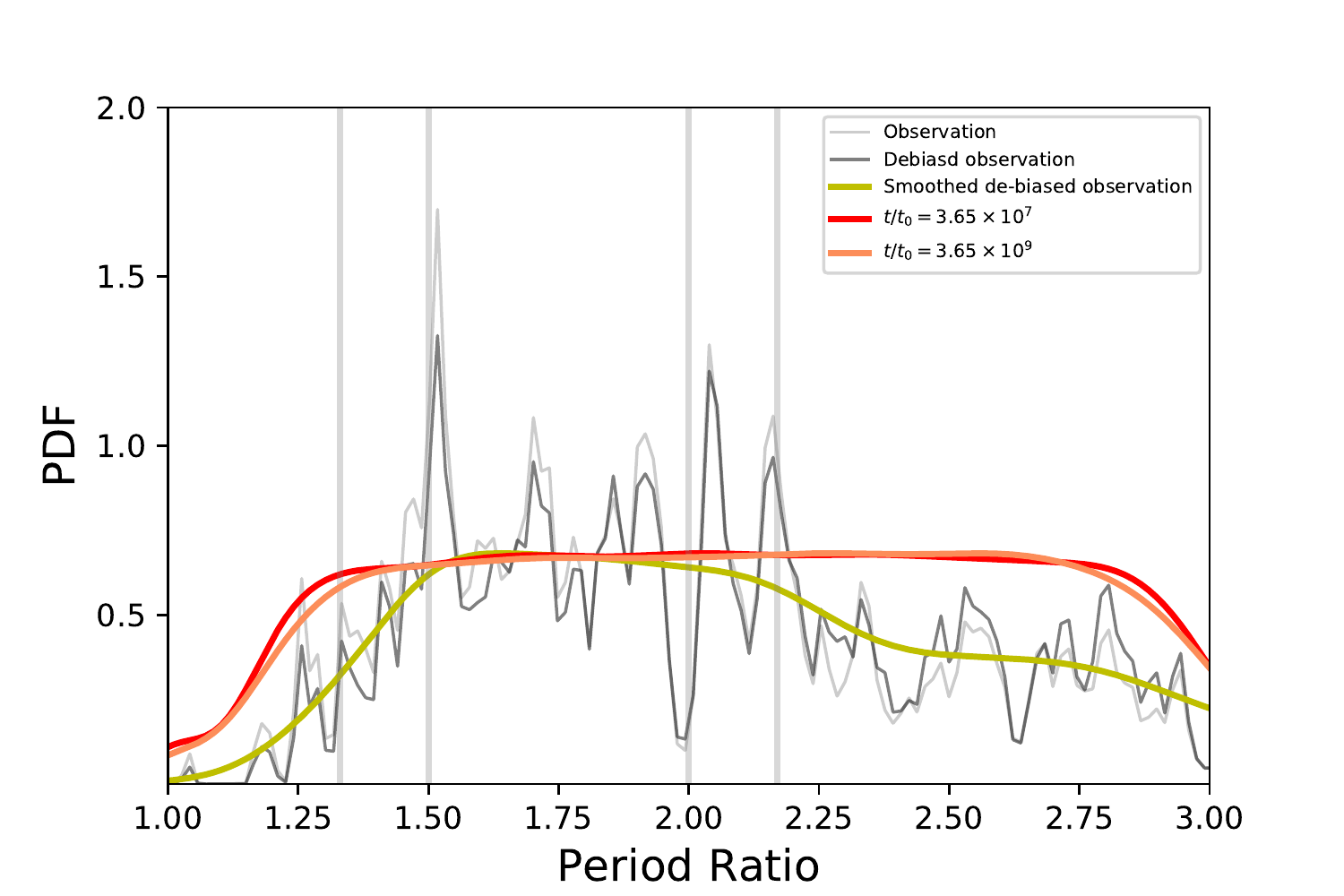}
\caption{Upper panel: the probability density function (PDF) of the period ratio for different kind of planetary systems described in Section \ref{sec:set-up} after $3.65\times10^7$ $t_0$. For a comparison, the PDF of all planetary systems are also shown (the red curve). Lower panel: The comparison between the observed and simulated period ratio distributions.  The PDF of the original period ratio distribution from \kepler\ observations is shown in gray.  PDF of the period ratio from de-biased \kepler\ observations is shown in light dark.  Yellow represents the smoothed PDF of the de-biased \kepler\ observations.  Red represents the re-normalized PDF of the period ratio from the simulation with evolution time of $3.65\times10^7$ $t_0$, while orange represents the re-normalized PDF of the period ratio of four-planet systems with an integration time of $3.65\times10^9$ $t_0$. The gray vertical lines indicate period ratios at 4:3, 3:2, 2:1 and 2.17.}f
\label{pdfb|a}
\end{figure}

The PDF of the observed period ratio and the de-biased period ratio distributions are shown in the lower panel of Figure \ref{pdfb|a}.  After de-biasing, the peaks near 3:2 and 2:1 MMRs persist, but they are not as significant as the original ones (especially for the peak near the 3:2 MMR) which are also seen in Figure 4 of \citet{brakensiek2016}.  We calculate the weight of each period ratio as the inverse of transiting probability of the outer planet given that the inner planet is transiting.  Additionally, we discuss another weighting scheme---which uses the inverse probability for both planets transiting the host star (rather than the conditional probability)---in the Appendix.

\subsubsection{Comparison of the PDF between observation and simulation}\label{sec:com}

In order to compare the observed \kepler\ period ratio distribution with our simulation (described in Section 2) results, we smooth the PDFs of both samples with suitable bandwidth.  The bandwidth is chosen as the smallest value that gives a unimodal distribution. The smoothed PDFs of the four kinds of planetary systems are shown in the upper panel of Figure \ref{pdfb|a}. For a comparison, the PDF of all planetary systems is also shown. We see that the shape of the PDF for the four-planet systems is very close to that of all planetary systems.  The comparison between the smoothed PDFs of all samples in our simulations and the observation is shown in the lower panel of Figure \ref{pdfb|a}. Note that the decrease of PDF at period ratio $>2.75$ is caused by the smoothing method and is not necessarily physical.  Moreover, while the period ratios in our simulations are distributed only between one and three, we can reasonably assume that period ratios larger than three will remain stable and that the final distribution will likely match the initial distribution for any simulated system in that regime.  We re-normalize the PDF of our simulation such that the largest value of the PDF of the observations and our simulation coincide.  (That is, we increase the height of the simulated distribution so that it matches the overall height of the observations.)  Differences in the two distributions following this modification should indicate period ratios where planets are under-represented relative to what dynamical stability would otherwise allow.

We find that the PDF of our simulation and the observations (the red and yellow curves in the lower panel of Figure \ref{pdfb|a}) roughly coincide between period ratios of 1.5 and 2.1.  For the deficit of planet pairs with period ratios between 1.1 and 1.5, the data show fewer systems than what our simulations suggest could survive.  However, given our limited integration time, there may be some residual instabilities that have not had time to manifest. Rather than continuing to integrate all planetary systems to a longer time, we simulate a set of four-planet systems and integrate them to $3.65\times10^9$ $t_0$ (100 times longer than the previous simulations).  We choose the four-planet systems for further integration because the shape of the PDF for four-planet systems roughly resembles the shape of the PDF for the whole samples. (see upper panel of Figure \ref{pdfb|a}).  The new simulations contain five hundred four-planet systems with the parameter distributions described in section 2. The re-normalized PDF of the new simulations is shown as the orange curve in the lower panel of Figure \ref{pdfb|a}.  We see that the shape of the new PDF changes very little when compared to that of the previous simulations (the red curve in the lower panel of Figure \ref{pdfb|a}).  Therefore, we suspect the shape of the PDF at period ratio $<1.5$ in the observation is not entirely due to instability, but may also be influenced by the initial eccentricity distribution, which we will discuss later.

For planet pairs with period ratio $> 2.1$, there is an obvious deficit in the observations when compared with the prediction of planet pairs that would otherwise survive given our simulations.  Since systems with period ratios this large should be stable for very long time (i.e., longer than the age of the universe), these results indicate planet pairs do not emerge from the protoplanetary disk with those period ratios to the same degree that they do with smaller period ratios, at least for systems like those observed by \kepler .  Thus, whatever formation or dynamical processes are ongoing while the protoplanetary disk is present, the frequency of planet pairs that are produced with period ratios between 2.1 and 3 is 30-50\% lower than the frequency of those produced between 1.5 and 2.1.


The sizable fraction of planet pairs that survive in the 1:1 MMR is at odds with the lack of observed planet pairs in those orbits.  This discrepancy likely indicates that planets either rarely form or are rarely driven into those configurations---if they did form, a large fraction would have survived.  It is possible that such planet pairs have been missed by the transit search algorithms, but the high signal-to-noise ratios of many of the \kepler\ detections makes this explanation difficult to justify in most cases.  (Though, we recommend revisiting the \kepler\ discoveries with this in mind.)

\subsubsection{Eccentricity of multi-planet systems when gas disk dissipates}

\begin{figure}
\vspace{0cm}\hspace{0cm}
\centering
\includegraphics[width=\columnwidth]{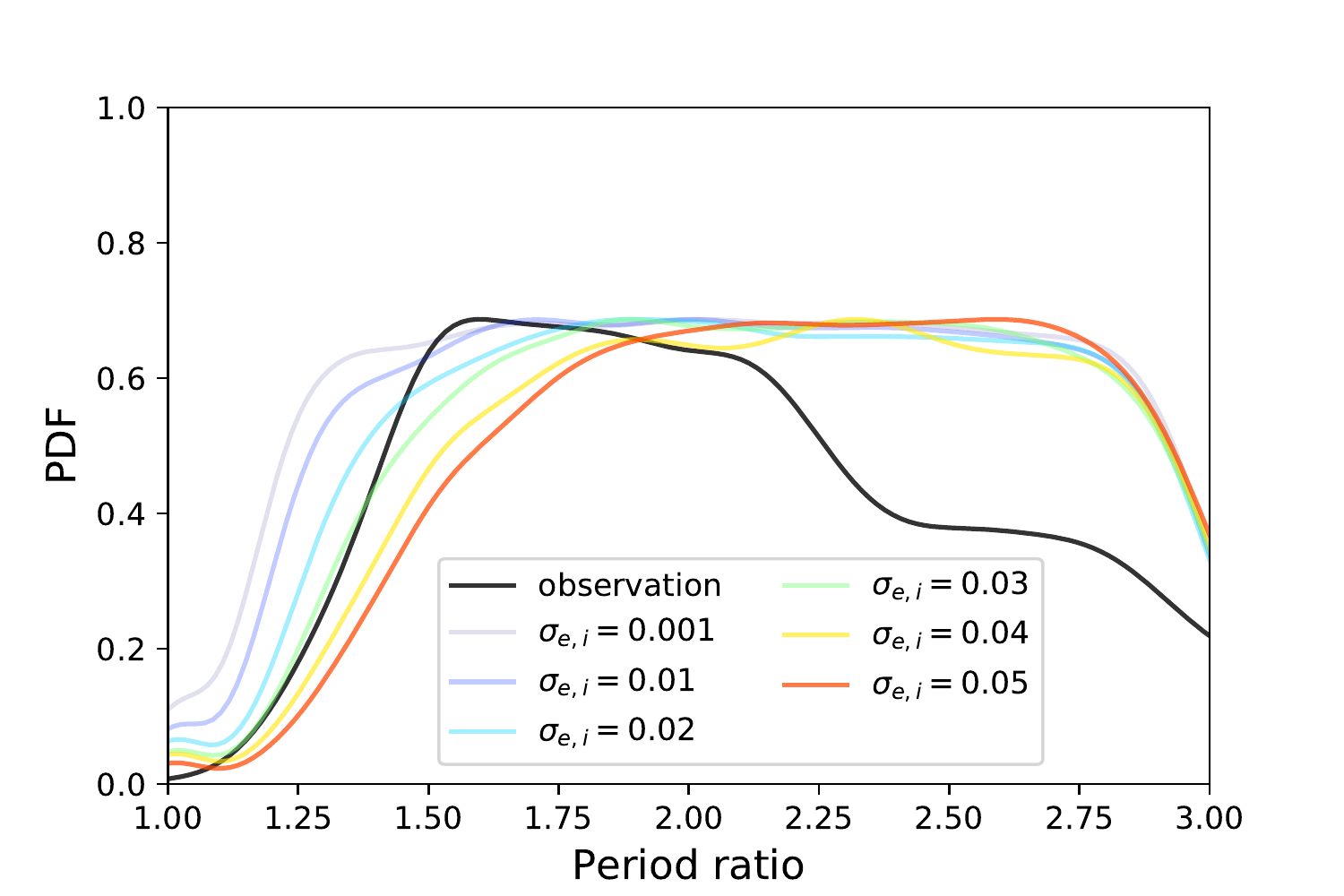}
\caption{The smoothed probability density function (PDF) of the period ratio for the observation (dark) and the re-normalized PDF of the period ratio from our simulations with different initial eccentricity and inclination distributions.  }
\label{comp_e}
\end{figure}

We showed that the observed period ratios between 1.1 and 1.5 can not be explained by the effects of instability with initial orbits that are nearly circular.  But We see from Figure \ref{pr_Rl_e} that orbital eccentricity drives more planet pairs with small period ratios into instability.  While the eccentricities of planets are likely to grow during the dynamical evolution following the dispersion of the gas disk, we can constrain the maximum initial eccentricity by comparing the shape of the PDFs between the observations and our simulations using different values of initial eccentricity.  We conduct a set of simulations with the same orbital parameters as those described in Section \ref{sec:set-up}, except for the eccentricity and the inclinations. The integration time is $3.65\times10^7$ $t_0$. The results are shown in Figure \ref{comp_e}.

We see from these simulations that systems with initial eccentricities and inclinations $\sigma_{e,i} \simeq 0.03$ are roughly consistent with the observed period ratios between 1.1 and 1.5 while larger values of initial eccentricity do not match the profile of the observed distribution.  Thus, the eccentricity and inclination distributions should have typical values $\sigma_{e,i} < 0.03$ when the gas disk dissipates. \citet{xie2016} proposed $e=0.04\pm0.04$ for multi-planet systems, which places an upper limit to the initial eccentricities around 0.04. Our prediction that $\sigma_{e,i} < 0.03$ is consistent with their limit.

\section{Conclusion}\label{sec:conclusion}

In this paper, we studied non-EMS multi-planet systems to investigate their stability and the evolution of their period ratio distribution.  In contrast to previous works, which assume the planets have equal mutual separation after the disk dissipates \citep{chambers1996,zhou2007,Smith2009,Obertas2017}, we begin with the premise that the orbital periods between adjacent planet pairs in multi-planet systems are uniformly distributed.  Thus, any differences between the observed distribution of period ratios and the results of our simulations are likely due to some physical process other than dynamical instability.

After an evolution time of $3.65\times10^7 \ t_0$, we find that surviving planet pairs with orbital period ratios $<1.1$ are protected by the 1:1 MMR (both in two-planet systems and systems with more than two planets).  These planets can be stable for $3.65\times10^9 \ t_0$ or longer whether in tadpole or horseshoe orbits.  Thus, the lack of co-orbital planet pairs in the observations indicates that either such planets are difficult to detect (which seems unlikely), or are rarely produced in planetary systems similar to those seen by \kepler .  If there was a viable mechanism to produce a large population of 1:1 MMR planet pairs, many would survive and should be seen.

For planets far from the 1:1 MMR, the lower limits of their stability timescales determined by $K_{\rm min}$ are consistent with what is predicted in EMS systems.  While planets in our simulations are not of equal mass, the differences between them are within one order-of-magnitude, our results should be largely unchanged as the Hill radius depends only weakly on planetary mass.  Of the statistical quantities we studied to characterize instability timescales, we find that $K_{\rm min}$ performs most consistently.

Our period ratio distribution shows a dip-peak asymmetry near first order MMRs, where more planets are on the far side of the resonance than near side.  We find that period ratios that are initially on the near side of these resonances are observed on the far side of the resonance more often due to their orbital evolution.  This result may partly explain the observed features near MMR in the \kepler\ data.  (Period ratios farther from the first order MMRs do not show such asymmetries in their orbital evolution.)  This deviation of the period ratio near MMR increases with planetary mass.  \citet{petrovich2013} proposed that in order to explain the observed asymmetric structure, the planetary mass should be in the range of 20 $-$ 100 $m_{\oplus}$.  However, the TTV-determined masses in \citet{Hadden2017} are too small to account for the dip-peak feature of the \kepler\ systems.  We also investigate the influence that eccentricity can have on the period ratio distribution and find that the dip-peak structure depends inversely upon the eccentricity of the planetary orbits---larger eccentricities show smaller asymmetry.  A non-zero initial eccentricity distribution with $\sigma_e=0.05$ is too large to produce the dip-peak structure.

Finally, we compare the probability density function of the de-biased period ratio distribution of the \kepler\ observation to our simulations.  We find that the general shape of the period ratio distribution less than $\sim2.1$ can be explained by dynamical instability of planetary systems with non-circular orbits with initial eccentricities $\lesssim 0.03$.  This same eccentricity preserves the asymmetry features near MMR while larger eccentricities simultaneously alters the resulting period ratio distribution removes the asymmetries.  (We note, however, that the asymmetries near MMR may not be caused by the mechanism we present here).  Local features near MMR and near 2.17 \citep{steffen2015} may require unique explanations.

We also find an obvious deficit of planet pairs with period ratios $\gtrsim 2.1$ in the \kepler\ data (the deficit is nearly 50\% of what would survive if they were initially present).  Thus, we suspect that planet pairs are either not formed as often with these period ratios, or if they are produced, that interactions with the gas disk may drive them to smaller period ratios.  For example, it may be that the initial distribution of period ratios is essentially flat, but that $\sim 25$\% of the planet pairs eventually converge to period ratios between 1.5 and 2.1---producing the two-plateaus shown in Figure \ref{pdfb|a}.

\kepler\ planetary systems are often portrayed as compact since planet pairs typically have small period ratios and orbit close to their host star.  However, the criteria for describing a system this way is ill defined.  Dynamical processes for planetary orbits are scale invariant, where resonance or other effects occur near certain period ratios regardless of the overall size of the system.  Only when some new physical scale enters the description is the invariance broken and the dynamics changed.  The results from \citet{Rice2018} indicate that dynamical effects related to instability are not markedly different between systems at 0.1 AU (where most \kepler\ planets are found) and at 1 AU where the solar system terrestrial planets are found---though more work on this issue is warranted.  The only scale where planetary system architecture is seen to change in the observations of \kepler\ planets is when the inner planet has an orbital period less than a few days \citep[$\sim 0.05$ AU][]{SteffenFarr2013,SteffenCoughlin2016}.  Moreover, period ratios observed in the solar system are similar to period ratios observed in most \kepler\ systems.  With the exception of the Jupiter/Mars ratio, solar system period ratios lie between 1.5 and 3 with the majority being less than 2.5.  Thus, unless the solar system is considered to be ``compact'' there is little to suggest that the typical \kepler\ planetary system should be so described.

This work shows that instability plays a significant role in sculpting planetary system architectures for period ratios less than 1.33 (see lower panel of Figure \ref{pdfb|a}).  These results suggest that a reasonable criterion for ``compactness'' could be that for a system to be considered compact, it must contain a planet pair with a period ratio less than this value.  For the \kepler\ multiplanet systems, this criterion would classify roughly 4\% of the systems as compact (or roughly 6\% of systems containing more than two planets---which may be more representative of multiplanet systems generally).

\section*{Acknowledgements}

We thank the anonymous referee who helped us to improve this paper. Thanks for useful discussion with Daniel C. Fabrycky, Ji-Wei Xie and Songhu Wang. This work is supported by the National Natural Science Foundation of China (grant No. 11503009, 11333002, 11673011, 11661161014), Technology of Space Telescope Detecting Exoplanet and Life supported by National Defense Science and Engineering Bureau civil spaceflight advanced research project (D030201), and China Scholarship Program.  JHS acknowledges support from the NASA Kepler Participating Scientist Program under grant number NNX16AK32G and the NASA Exoplanet Research Program under grant number NNX17AK94G.

\appendix

\section{Period ratio de-biasing}

\subsection{Two kinds of weight calculation}
In Section \ref{subsec:pdf}, we discuss the probability density function (PDF) of the de-biased period ratios.  The weight of each period ratio is assumed to be the inverse of the transiting probability of the outer planet given that the inner planet is transiting.  This is the method used in \citet{steffen2015}.  We record the PDF calculated this way as $\mathcal{F}(out|in)$.

\begin{figure}
\vspace{0cm}\hspace{0cm}
\centering
\includegraphics[width=\columnwidth]{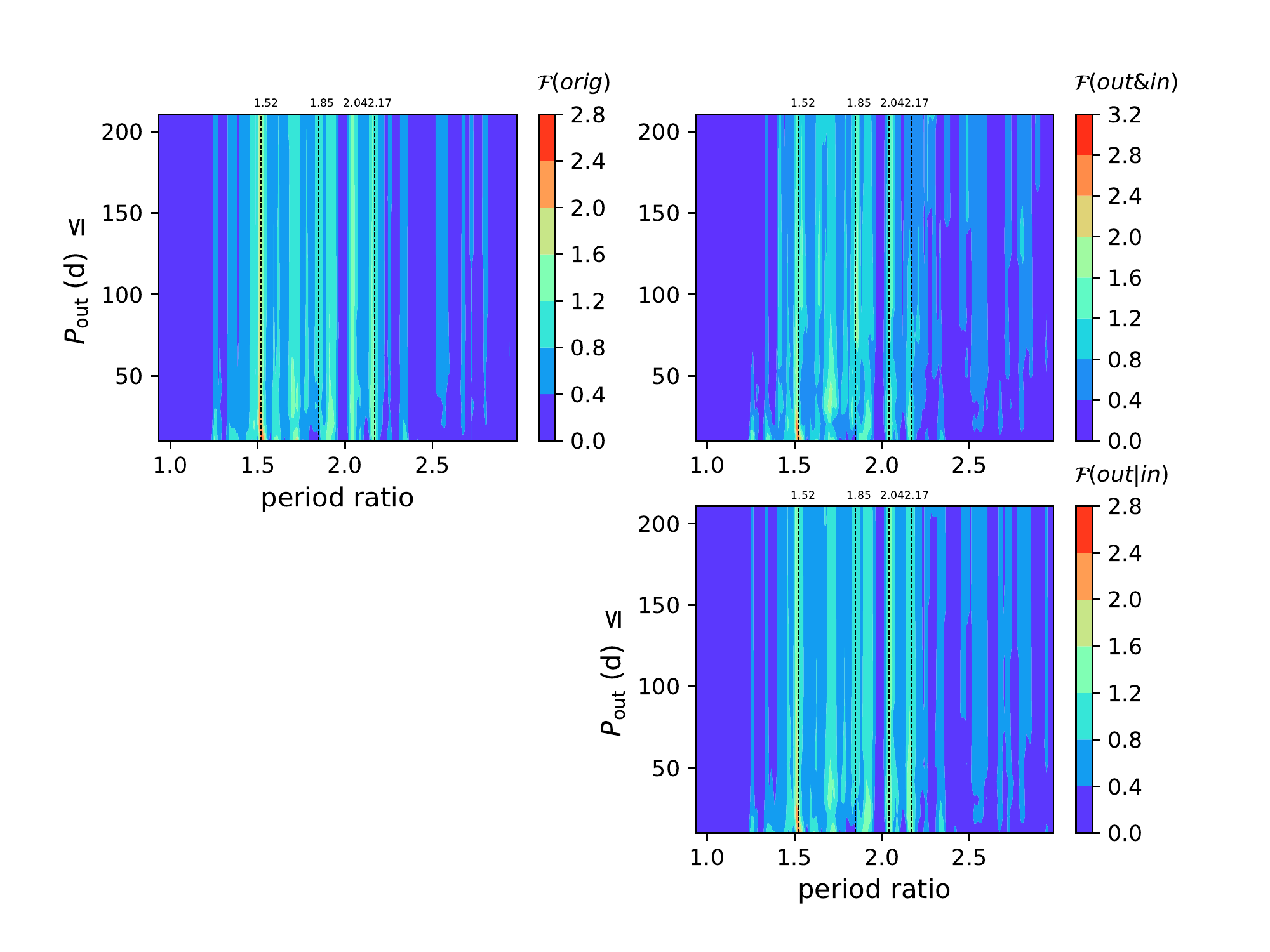}
\caption{The probability density function (PDF) of the period ratio as a function of outer orbital period $P_{\rm out}$ of the planet pair.  The samples used here are the same as Figure \ref{observation}.  The y axis means we include planet pairs with $P_{\rm out}$ smaller than one specific value on the y axis.  The upper left panel represents the original PDF $\mathcal{F}(orig)$ of period ratios, and the upper right panel represents the de-biased PDF of period ratios $\mathcal{F}(out\&in)$, where the weight is calculated as the inverse of the probability when both planets are transiting the host star.  The lower right panel represents the de-biased PDF of period ratios $\mathcal{F}(out|in)$, where the weight is calculated as the inverse of the transiting probability given that the inner planet is transiting.  The dashed vertical lines show the period ratios at 1.52, 1.85, 2.04 and 2.17.}
\label{figa1}
\end{figure}

Another weighting method could be the inverse of the probability when both planets are transiting the host star directly, $\mathcal{F}(out\&in)$.  It is not obvious which of these approaches is correct.  The assumption in \citet{steffen2015} is that you would not detect a planet pair if you had not detected the inner planet in that pair---hence their use of $\mathcal{F}(out|in)$.  Either way, we show the PDF of the original $\mathcal{F}(orig)$ and the two kinds of de-biased period ratios as a function of orbital period in Figure \ref{figa1}, respectively.  The transiting probabilities are calculated with CORBITS described in \citet{brakensiek2016}.

Compared with $\mathcal{F}(orig)$, $\mathcal{F}(out|in)$ increases at larger period ratios and decreases at smaller period ratios, while $\mathcal{F}(out\&in)$ shows more obvious change with $P_{\rm out}$. 
The peaks in the PDF at 1.52 and 2.04 appear in all three period ratio distributions.  However, when we consider only orbital periods $<20$ d, there is no significant peak at 1.85.  As orbital period increases, the peak at 1.85 appears, but is not as significant as the peak at 1.52 and 2.04, especially for $\mathcal{F}(orig)$ and $\mathcal{F}(out|in)$.  

Another interesting peak is 2.17.  It exists in $\mathcal{F}(orig)$ and $\mathcal{F}(out|in)$ for all orbital periods, but for $\mathcal{F}(out\&in)$, the peak at 2.17 disappears once we include planet pairs with orbital period $>$ 130 d.  We checked the samples with orbital period ratios near 2.17 and find that they mainly constitute of planet pairs with orbital periods between 10-20 d.  Hence, in the calculation of $\mathcal{F}(out\&in)$, the inclusion of planet pairs with long orbital periods increases the weight of other period ratios and simultaneously reduces the weight of the period ratio at 2.17.  Thus, it may be that the process that creates the feature at 2.17 is something that occurs only in the innermost parts of the protoplanetary disk.

\subsection{Influence of mutual inclination between planet pairs}

\begin{figure}
\vspace{0cm}\hspace{0cm}
\centering
\includegraphics[width=\columnwidth]{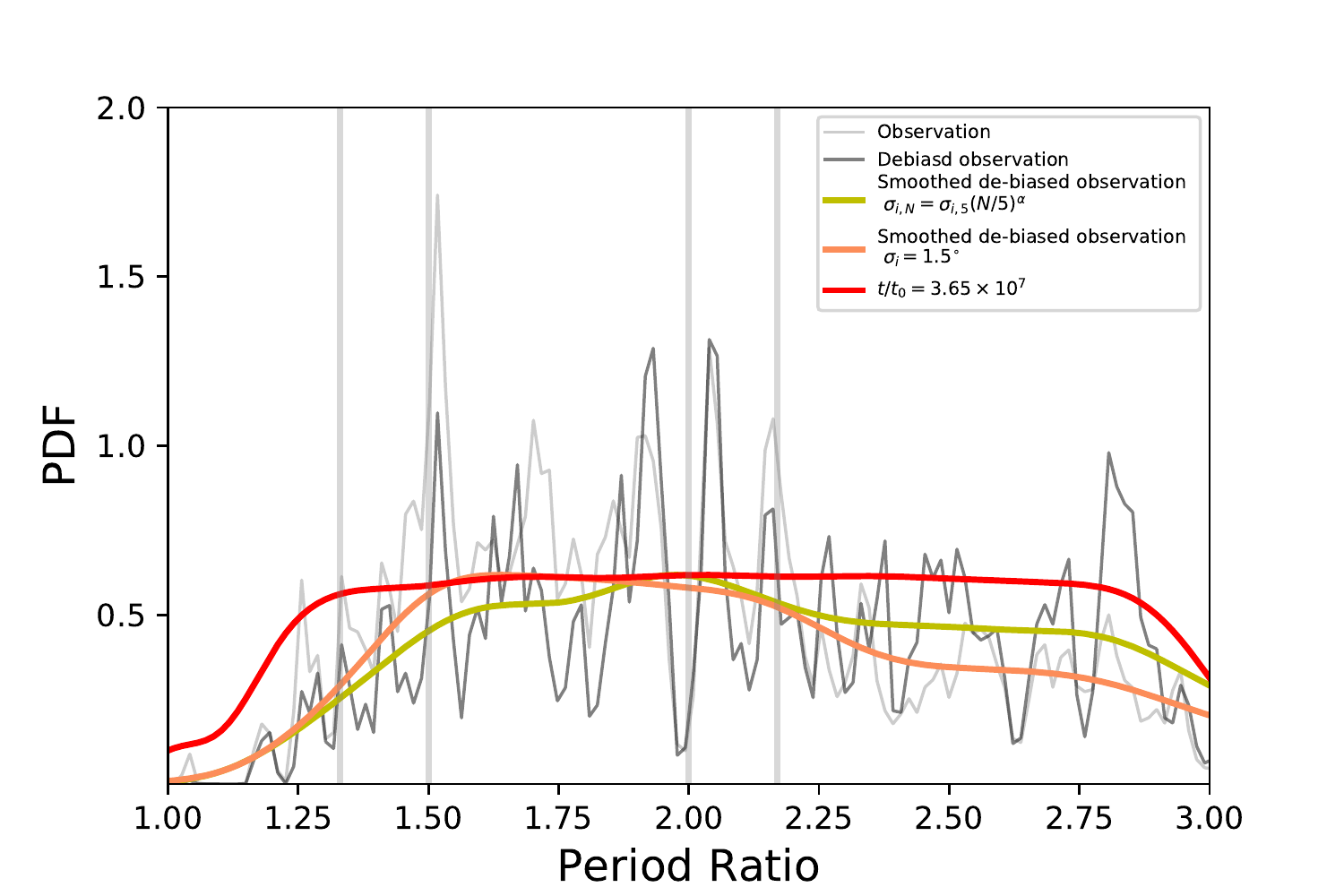}
\caption{The comparison of the PDF between the \kepler\ observation and our simulations. The observed period ratio is de-biased assuming that the inclination dispersion of each planet is $\sigma_{i,N}=\sigma_{i,5}(N/5)^{\alpha}$, where $N$ is the number of planets in a given planetary system, $\sigma_{i,5}\approx 0.8^{\circ}$, $\alpha=-4$. The smoothed PDF of the de-biased observation is shown as the yellow curve. For a comparison, the smoothed PDF of the co-planar case (where the mutual inclination between planet pairs is assumed to be around $1.5^{\circ}$) is shown as the orange curve. The simulated PDF with an integration time of $3.65\times10^7$ $t_0$ is shown as the red curve.}
\label{pdfi}
\end{figure}

In section \ref{subsec:pdf}, we assumed that the mutual inclination of planets in a system are Rayleigh distributed with $\sigma$ $\sim$ $1.5^{\circ}$ (noted as the co-planar case).  However, \citet{Zhu:2018} proposed that the dispersion of planetary inclinations within a given system is a function of its number of planets $N$, i.e., $\sigma_{i,N}=\sigma_{i,5}(N/5)^{\alpha}$, where $\sigma_{i,5}\approx 0.8^{\circ}$, $-4<\alpha<-2$. Based on this inclination distribution function, We recalculate the transiting probability of each planet pair assuming an extreme case where $\alpha=-4$ (noted as the inclined case). The weight of each period ratio is calculated using the method described in section \ref{subsec:pdf}. The PDF for the inclined case, the co-planar case and our simulation are shown in Figure \ref{pdfi}. Compared to the co-planar case, the PDF at period ratio $>2$ for the inclined case increases, while the PDF at period ratio $<2$ decreases. It is because that most of planet pairs with period ratio $>2$ are from two-planet systems. If we assume a larger mutual inclination for two-planet systems than planetary systems with higher multiplicity, the weight of period ratio $>2$ will increase, which leads to the increase of PDF at period ratio $>2$.  Nevertheless, there is still deficit of planet pairs in the observation at period ratio $>2.1$.

\bibliographystyle{mnras}
\bibliography{ms.bib}

\label{lastpage}
\end{document}